\documentclass[conference]{IEEEtran}
\usepackage{aliascnt}
\usepackage{cite}
\usepackage{wrapfig}
\usepackage{gensymb}

\IEEEoverridecommandlockouts
\newenvironment{talign}
 {\align}
 {\endalign}
\usepackage{multirow}
\usepackage{enumitem}
\usepackage{pifont}
\usepackage{amsmath,amssymb,amsfonts,amsthm}
\usepackage{algorithmic}
\usepackage[ruled,linesnumbered]{algorithm2e}
\usepackage{graphicx}
\usepackage{textcomp}
\usepackage{caption}
\usepackage[hidelinks]{hyperref}
\usepackage{float}
\usepackage[utf8]{inputenc}
\usepackage{tabularx}
\usepackage{xcolor}
\usepackage{booktabs}
\usepackage{cases}
\usepackage{subfigure}
\usepackage{bm}
\def\BibTeX{{\rm B\kern-.05em{\sc i\kern-.025em b}\kern-.08em
    T\kern-.1667em\lower.7ex\hbox{E}\kern-.125emX}}
    
\usepackage{graphicx}
\usepackage{mathrsfs}
\usepackage{bm}

\newtheoremstyle{slanted}
{0em plus 0em minus 0em}
  {0em plus 0em minus 0em}
  {\em}
  {}
  {\bfseries}
  {.}
  { }
  {}
\theoremstyle{slanted}

\theoremstyle{slanted}

\theoremstyle{slanted}

\theoremstyle{slanted}

\theoremstyle{slanted}

\theoremstyle{slanted}

\theoremstyle{slanted}

\usepackage{enumitem}

\title{QuHE: Optimizing Utility-Cost in Quantum Key Distribution and Homomorphic Encryption Enabled Secure Edge Computing Networks}

\author{\IEEEauthorblockN{Liangxin Qian, Yang Li, and Jun Zhao\thanks{Corresponding author: Jun Zhao.\newline \indent Liangxin Qian and Yang Li are both PhD students supervised by Jun Zhao. Jun Zhao's ORCID: 0000-0002-3004-7091 }\\}
\IEEEauthorblockA{{College of Computing and Data Science} \\
{Nanyang Technological University, Singapore}\\
\{qian0080, yang048\}@e.ntu.edu.sg, junzhao@ntu.edu.sg\vspace{-15pt}}
}

\begin{document}
\maketitle
 \thispagestyle{plain}
\pagestyle{plain}
\begin{abstract}
Ensuring secure and efficient data processing in mobile edge computing (MEC) systems is a critical challenge. While quantum key distribution (QKD) offers unconditionally secure key exchange and homomorphic encryption (HE) enables privacy-preserving data processing, existing research fails to address the comprehensive trade-offs among QKD utility, HE security, and system costs. This paper proposes a novel framework integrating QKD, transciphering, and HE for secure and efficient MEC. QKD distributes symmetric keys, transciphering bridges symmetric encryption, and HE processes encrypted data at the server. We formulate an optimization problem balancing QKD utility, HE security, processing and wireless transmission costs. However, the formulated optimization is non-convex and NP-hard. To solve it efficiently, we propose the \underline{Qu}antum-enhanced \underline{H}omomorphic \underline{E}ncryption resource allocation (QuHE) algorithm. Theoretical analysis proves the proposed QuHE algorithm's convergence and optimality, and simulations demonstrate its effectiveness across multiple performance metrics.
\end{abstract}


\section{Introduction}
\subsection{Research Background}
The rapid development of mobile edge computing (MEC) and wireless communications has fueled demand for secure and efficient data processing at the network edge \cite{shan2022computing,jiao2023deep}. Internet of Things (IoT) applications rely on real-time data analytics to deliver high performance and privacy \cite{chen2024privacy,zhao2021joint}. In these dynamic and resource-constrained environments, ensuring data security without compromising computational efficiency remains a significant challenge.

Homomorphic encryption (HE) has emerged as a transformative solution for secure data processing in MEC systems~\cite{mohammed2022fully}. Unlike traditional cryptographic methods, HE enables computations directly on encrypted data, ensuring data confidentiality throughout the entire computation process. This makes HE well-suited for privacy-preserving edge computing applications where sensitive data must remain secure during analysis and processing. However, the reliance of HE on asymmetric cryptography often results in computational overhead and complex key management, posing challenges for distributed MEC systems with limited resources.

As a potential technique, quantum key distribution (QKD) offers an innovative approach to secure key exchange by leveraging the principles of quantum mechanics, e.g., superposition and entanglement \cite{ding2018enhancing}. Unlike classical methods, QKD provides unconditional security against eavesdropping by detecting any interception of quantum states during transmission. This capability enables the generation and distribution of symmetric keys with unmatched reliability. Integrating QKD into HE-based MEC systems can enhance security by replacing computationally intensive asymmetric cryptographic operations with efficient, symmetric key-based encryption.

\subsection{Motivation}
As MEC systems continue to grow in scale and complexity, ensuring robust security and efficient resource utilization has become increasingly critical. While QKD networks have been extensively studied for their ability to provide secure key exchange, existing research \cite{vardoyan2023quantum, pouryousef2023quantum, lee2024quantum, kar2024convexification, herrmann2023quantum} has predominantly focused on optimizing QKD network utility without addressing its integration with HE systems. Specifically, these works neglect the importance of incorporating the minimum security level in HE and fail to account for the substantial processing and wireless transmission costs that arise in MEC systems.

Similarly, studies on QKD-enabled HE systems \cite{ding2018enhancing,lemons2023extending,diovu2019enhancing} primarily emphasize the protocol design or theoretical framework, overlooking the challenges of balancing security and resource efficiency. Besides, research on resource allocation in HE systems \cite{shan2022computing,mohammed2022fully,sheela2024secure,chen2024privacy} focuses mainly on optimizing processing costs without considering the advantages of QKD or the minimum security requirements in HE.

To date, no existing research addresses the trade-off between QKD network utility, minimum security level in HE, processing, and wireless transmission costs. This gap highlights the need for a holistic optimization framework that balances these interconnected factors. Motivated by this, our study proposes an innovative approach to integrate QKD with HE and transciphering in MEC systems, aiming to optimize the trade-off between QKD utility-HE security and resource efficiency while ensuring reliable and effective system performance.

\subsection{Studied Problem and Contribution}
This study focuses on designing and optimizing a QKD-enhanced FHE-based edge computing system. The key problem is to balance the trade-off between quantum network utility, the minimal security level, and communication and computation costs. To address this, the system's optimization problem is formulated to maximize quantum network utility and security levels while minimizing communication and computation costs. This non-convex and NP-hard optimization problem is tackled using the proposed QuHE algorithm. The main contributions of this paper are summarized as follows:
\begin{itemize}
    \item We design a novel edge computing system that integrates QKD for secure symmetric key distribution with FHE for privacy-preserving computation. The system ensures robust security, efficient key management, and encrypted data processing.
    \item We formulate a non-convex optimization problem to balance QKD network utility, minimum security level, and communication and computation costs. The formulation reflects the trade-offs in QKD-enhanced FHE systems.
    \item We propose the \underline{Qu}antum-enhanced \underline{H}omomorphic \underline{E}ncryption resource allocation algorithm (QuHE) to solve the NP-hard optimization problem efficiently. The algorithm combines heuristic search methods and theoretical insights to achieve reliable solutions.
    \item We analyze the convergence, solution optimality, and complexity of the proposed QuHE algorithm, providing theoretical guarantees for its performance.
    \item Through extensive simulations, we verify the effectiveness and reliability of the QuHE algorithm under various resource configurations and demonstrate its superiority over baseline methods in terms of energy efficiency, delay, and security.
\end{itemize}
The remainder of this paper is structured as follows. Section \ref{sec_relatedwork} reviews the related work, highlighting the novelty of our study compared to existing work. The system model and the optimization problem formulation are detailed in Sections \ref{sec_systemmodel} and \ref{sec_optiproblem}, respectively. In Section \ref{sec_quhe}, we introduce the proposed QuHE algorithm to address the formulated problem and provide an in-depth analysis of its convergence, solution optimality, and computational complexity. Numerical simulations and performance evaluations are presented in Section \ref{sec_simulation}. Finally, Section \ref{sec_conclusion} concludes the paper.

\section{Related Work} \label{sec_relatedwork}
In this section, we present the related work in the research of the utility of QKD networks, QKD-enabled HE systems, and resource allocation in HE systems. Then, the novelty of our paper compared to the related work is discussed.
\subsection{Utility in QKD Networks}
QKD networks have gained attention for enabling secure communication, with utility maximization emerging as a critical focus. Vardoyan \textit{et al.} \cite{vardoyan2023quantum} extended classical network utility maximization to quantum networks, optimizing resource allocation based on entanglement measures and exploring fidelity-rate trade-offs. Pouryousef \textit{et al.} \cite{pouryousef2023quantum} developed a quantum network planning framework to maximize utility by efficiently deploying quantum hardware. Lee \textit{et al.} \cite{lee2024quantum} introduced a benchmarking framework to evaluate quantum networks' social and economic value. Kar and Wehner \cite{kar2024convexification} formulated a convex optimization approach for quantum network utility maximization, addressing heterogeneous network routes. Herrmann \textit{et al.} \cite{herrmann2023quantum} proposed quantum utility as a practical measure of quantum systems' advantages.

\subsection{QKD-Enabled Homomorphic Encryption Systems}
The integration of QKD with HE has demonstrated significant potential for enhancing security in distributed systems. Ding \textit{et al.} \cite{ding2018enhancing} proposed a QKD-enhanced HE framework for securing multi-agent networked control systems, leveraging the randomness of quantum keys and symmetric encryption to reduce computational overhead while maintaining strong security. Lemons \textit{et al.} \cite{lemons2023extending} addressed the challenge of extending QKD networks over long distances by combining QKD with homomorphic key-switching, enabling secure multi-party key sharing through relay nodes with lattice-based cryptographic implementations demonstrating feasibility. In smart grids, Diovu and Agee \cite{diovu2019enhancing} proposed a cloud-based advanced metering infrastructure fortified by QKD, ensuring data confidentiality and integrity while maintaining scalability through lightweight protocols. These works highlight the advantages of QKD in improving security and key management across various domains, forming a strong basis for exploring its application in secure edge computing networks.
\begin{table*}[htbp]
\centering
\caption{Comparison of related work and this paper.}
\label{tab_comparison}
\begin{tabular}{|c|c|c|c|c|c|c|c|}
\hline
\textbf{Paper} & \textbf{QKD} & \textbf{QKD utility} & \textbf{HE} & \textbf{Minimum security level} & \textbf{Processing cost} & \textbf{Wireless} & \textbf{Transmission cost} \\ \hline
Vardoyan \textit{et al.} \cite{vardoyan2023quantum} & \checkmark & \checkmark & $\times$ & $\times$ & $\times$ & $\times$ & $\times$ \\ \hline
Pouryousef \textit{et al.} \cite{pouryousef2023quantum} & \checkmark & \checkmark & $\times$ & $\times$ & $\times$ & $\times$ & $\times$ \\ \hline
Lee \textit{et al.} \cite{lee2024quantum} & \checkmark & \checkmark & $\times$ & $\times$ & $\times$ & $\times$ & $\times$ \\ \hline
Ding \textit{et al.} \cite{ding2018enhancing} & \checkmark & $\times$ & \checkmark & $\times$ & $\times$ & $\times$ & $\times$ \\ \hline
Lemons \textit{et al.} \cite{lemons2023extending} & \checkmark & $\times$ & \checkmark & $\times$ & $\times$ & $\times$ & $\times$ \\ \hline
Shan \textit{et al.} \cite{shan2022computing} & $\times$ & $\times$ & \checkmark & $\times$ & \checkmark & $\times$ & $\times$ \\ \hline
Mohammed \textit{et al.} \cite{mohammed2022fully} & $\times$ & $\times$ & \checkmark & $\times$ & \checkmark & $\times$ & $\times$ \\ \hline
Sheela \textit{et al.} \cite{sheela2024secure} & $\times$ & $\times$ & \checkmark & $\times$ & \checkmark & \checkmark & $\times$ \\ \hline
Chen \textit{et al.} \cite{chen2024privacy} & $\times$ & $\times$ & \checkmark & $\times$ & $\times$ & \checkmark & $\times$ \\ \hline
This Paper & \checkmark & \checkmark & \checkmark & \checkmark & \checkmark & \checkmark & \checkmark \\ \hline
\end{tabular}
\end{table*}
\subsection{Resource Allocation in Homomorphic Encryption Systems}
HE has been effectively combined with resource allocation strategies to address challenges related to data privacy, system delay, and computational efficiency in various domains. Shan \textit{et al.} \cite{shan2022computing} proposed a privacy-preserving resource allocation strategy in edge computing systems using a partially observable Markov decision process and privacy entropy. Their method reduces system energy consumption and enhances security during data distribution and transmission. Mohammed \textit{et al.} \cite{mohammed2022fully} tackled resource allocation in vehicular fog cloud networks by introducing a cost-efficient and secure system leveraging FHE. Their framework addresses mobility and offloading costs while ensuring task deadlines are met, achieving significant cost optimization. Sheela \textit{et al.} \cite{sheela2024secure} integrated HE with reinforcement learning (RL) in wireless sensor networks (WSNs), enabling secure training and optimization of global models by performing computations on encrypted data. The framework also incorporates quantum-safe cryptographic techniques, offering robust security against quantum threats. Chen \textit{et al.} \cite{chen2024privacy} presented a privacy-preserving double auction mechanism for resource allocation in satellite MEC. By combining HE with garbled circuits and leveraging dynamic programming, their solution ensures security and privacy in auction-based resource allocation while maintaining efficiency. These works highlight the potential of integrating HE with advanced optimization techniques to achieve secure and efficient resource allocation across diverse applications.

\subsection{Novelty of Our Paper}
This paper is the first to study the integration of QKD into HE-enabled MEC systems with a focus on optimizing the trade-off between utility and costs. While existing studies explore QKD or HE in isolation, no prior work addresses the combined system or analyzes this specific trade-off. We utilize the QKD network to securely distribute symmetric keys from the key center to client nodes, enabling secure encryption and data transmission for further processing on the server side. The utility is defined as a combination of QKD network utility and the minimal security level of HE, while the costs encompass delay and energy consumption in encryption, wireless transmission, and server processing. We compare the related work and this paper in Table \ref{tab_comparison}.

\section{System Model and Parameter Description}\label{sec_systemmodel}
\begin{figure}[tbp]
\centering
\includegraphics[width=0.45\textwidth]{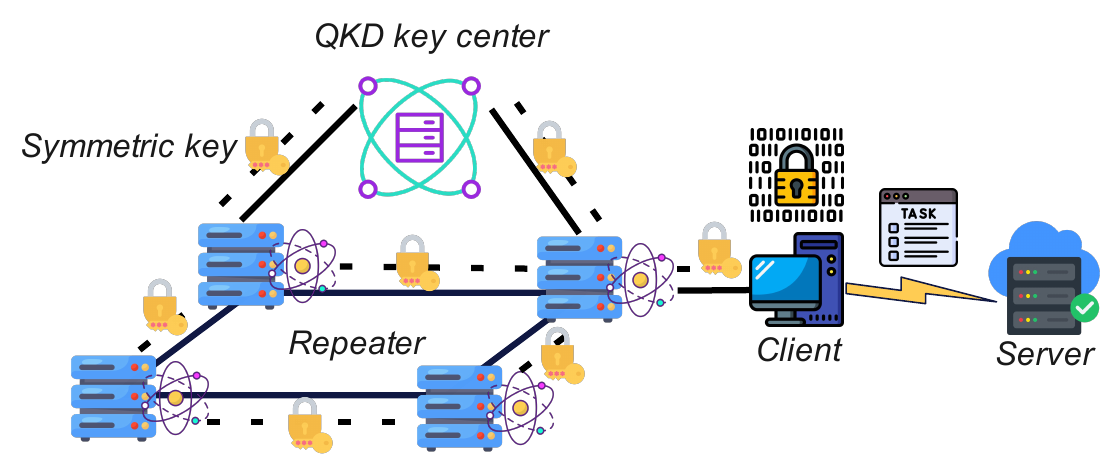}
\vspace{-6pt}
\caption{System model.}
\label{fig_system_model}
\end{figure}

In this section, we first give an overview of our studied system and then illustrate detailed parameters and metrics. Some important notations are shown in Table \ref{tab_notation}. The system model is shown in Fig. \ref{fig_system_model}.
\begin{table}[tbp]
\setlength{\abovecaptionskip}{5pt}
\setlength{\belowcaptionskip}{10pt}
\caption{Important notation.}\vspace{-3pt}\label{tab_notation}
\centering
\setlength{\tabcolsep}{0pt}
\small
\renewcommand{\arraystretch}{1.1}
\begin{tabular}{m{5em}  m{7cm}}
    \toprule                                   
    \textbf{Notation}  & \textbf{Description} \\
    \midrule
    $\mathcal{N}$ & The set of all route and client nodes ($n \in\{1,...,N\}$)\\ 
    $\mathcal{L}$ & The set of all links ($l \in\{1,...,L\}$)\\
    $w_l$ & The Werner parameter of the $l$-th link\\
    $\phi_n$ & The entanglement rate allocated to the $n$-th route\\
    $\lambda_n$ & The polynomial degree of client node $n$\\
    $p_{n}$ & The transmit power of client node $n$\\
    $b_n$ & The allocated bandwidth between client node $n$ and the server\\
    $r_n$ & The transmission rate from client node $n$ to the server\\
    $f_n^{(c)}$ & The available computing capacity of client node $n$\\
    $f_n^{(s)}$ & The computational capacity of the server allocated for client node $n$\\
    $f^{(eval)}(\lambda_n)$ & The CPU cycles needed to evaluate per sample with the polynomial modulus $\lambda_n$ \\
    $f^{(cmp)}(\lambda_n)$ & The CPU cycles needed to compute per sample with the polynomial modulus $\lambda_n$ \\
    $f^{(msl)}(\lambda_n)$ & The minimum security level of client node $n$\\
    \bottomrule
\end{tabular}
\vspace{-10pt}
\end{table}
\subsection{System Overview}
This section outlines the detailed process of integrating QKD with the CKKS homomorphic encryption scheme \cite{li2024privtuner} to achieve secure edge computing. We consider the uplink from the client nodes to the server in this study.
\subsubsection{Key Distribution via QKD} QKD is utilized to securely generate and distribute symmetric keys between a key center and client nodes. Unlike public key cryptography, which is vulnerable to quantum attacks, QKD guarantees unconditional security by enabling symmetric key exchanges over quantum channels. These symmetric keys form the foundation for secure operations in the system.
\subsubsection{Data Encryption Using Homomorphic Encryption}
Once the symmetric key $k_{qkd}$ is securely shared with the client node via QKD, one symmetric encryption (e.g., stream ciphers like ChaCha20 \cite{semenov2020implementation}) is performed with the received symmetric key, the plaintext data $m_p$ at the client node is encrypted into a ciphertext $c$ as
\begin{talign}
    c = \text{E}_{k_{qkd}}(m_p),
\end{talign}
where ``$\text{E}$'' means the symmetric encryption operation. Then, the client node runs the key generation algorithm for HE:
\begin{talign}
    \text{KeyGen}(\lambda, q) \rightarrow (\text{pk}, \text{sk}),
\end{talign}
where ``\text{KeyGen}'' is the key generation function, $q$ is a coefficient modulus, ``\text{pk}'' is the public key, and ``\text{sk}'' is the secret key. The client node also encrypts $k_{qkd}$ with the HE algorithm using the public key \text{pk}, i.e., $\text{Enc}(k_{qkd})$. 
\subsubsection{Encrypted Data Transmission}
The client node transmits the resulting ciphertexts $c$ and $\text{Enc}(k_{qkd})$ via wireless communication to the nearby server node for further operations. This process ensures that the data remains confidential during transit and at rest on the server.
\subsubsection{Transciphering at the Server and Encrypted Data Processing}
The server node first computes $\text{Enc}(c)$ for the received ciphertext $c$. Then the server homomorphically evaluated $\text{E}^{-1}$ over $\text{Enc}(c)$ and $\text{Enc}(k_{qkd})$, securely obtaining $\text{Enc}(m_p)$ \cite{cho2021transciphering}. Then, the server node performs computations directly on it using the homomorphic properties of CKKS. This phase eliminates the need to decrypt the data, preserving confidentiality throughout the computational process.
By doing so, the client no longer needs to perform the high overhead HE encryption, but leaves it to the server. This also decreases the transmission overhead in Phase 3).

\subsection{Quantum Key Distribution Phase Analysis}
Consider a QKD network with $L$ links and $N$ routes. Let $\mathcal{L}$ and $\mathcal{N}$ denote the sets of links and routes, respectively, defined as $\mathcal{L}:= \{1,2,\cdots,L\}$ and $\mathcal{N}:= \{1,2,\cdots,N\}$. The indices $l\in\mathcal{L}$ and $n\in\mathcal{N}$ are used to represent the $l$-th link and $n$-th route, respectively. For brevity, $n$ also denotes the index of the $n$-th client node, and the destination node of the $n$-th route is the $n$-th client node. Assume that there is one server node equipped with sufficient computation resources. There is one central key center, and it performs QKD service to deliver the secret keys to each client node via optical fibers. 

\subsubsection{QKD Network Utility} We use $w_l$ to denote the Werner parameter of $l$-th link. Define $\bm{w}:=[w_l]|_{l \in \mathcal{L}}$. The capacity of the $l$-th link, given a fixed $w_l$, is expressed as:
\begin{talign}
    c_l = \beta_l (1 - w_l),
\end{talign}
where $\beta_l$ is $3\kappa_l \eta_l /(2T_l)$. $\kappa_l$ is the inefficiency factor of the $l$-th link in optical fibers, excluding photon loss. $\eta_l$ is the transmissivity from one end of the $l$-th link to its midpoint. $T_l$ is the time the $l$-th link generates entanglement pairs. Next, we show how to formulate the QKD network utility.

We consider performing QKD using specific entanglement pairs of nodes within the quantum network. The secret key fraction, denoted as $F_{skf}(w)$, serves as the key measure of entanglement, where $w$ is the Werner parameter. Here, the subscript ``skf'' identifies the secret key fraction. The expression of $F_{skf}(w)$ is 
\begin{talign}
    &F_{skf}(w) \nonumber \\
    &= \max\left(0,1+(1+w)\log_2(\frac{1+w}{2})+(1-w)\log_2(\frac{1-w}{2})\right),
\end{talign}
Let $\varpi_n$ represent the end-to-end Werner parameter for the $n$-th route. The calculation for $\varpi_n$ is given by:
\begin{talign}
    \varpi_n = \prod_{l=1}^L w_l^{a_{ln}},
\end{talign}
where $a_{ln}$ is a binary variable that indicates whether the $l$-th link is part of the $n$-th route ($a_{ln} = 1$) or not ($a_{ln} = 0$). We define $\bm{A}:=[a_{ln}]|_{l \in \mathcal{L}, n \in \mathcal{N}}$ as the matrix describing the link-route relationship. The rate allocated to the $n$-th route is denoted by $\phi_n$, where $n\in \mathcal{N}$, and the vector of all allocated rates is defined as $\bm{\phi}:=[\phi_n]|_{n \in \mathcal{N}}$. Following the utility modeling framework presented in \cite{kar2024convexification}, the QKD network utility can be expressed as:
\begin{talign}
    U_{qkd} = \prod_{n=1}^N \phi_n F_{skf}(\varpi_n).
\end{talign}

\subsubsection{Reasons to Omit Costs in the QKD Network} 
The cost of QKD infrastructure is treated as a fixed, upfront investment and is assumed to be negligible for this study. Practical deployments often consider QKD infrastructure pre-installed. This work focuses on optimizing operational costs and delays related to client operations, uplink transmission, and server computations, ensuring a utility-centric approach centered on actionable system parameters. In the following part, we study the cost during the encryption phase of client nodes.
\subsection{Encryption Phase Analysis}
After receiving symmetric keys from the key center, each client node performs symmetric encryption tasks. The $n$-th client node corresponds to the $n$-th route's destination and processes natural language processing (NLP) tasks. The coefficient moduli $\bm{q}$ are fixed as large values to ensure sufficient arithmetic depth in FHE, while the polynomial degree $\bm{\lambda}:= [\lambda_n]|_{n \in \mathcal{N}}$ is optimized. 
\subsubsection{Encryption Delay} 
Define that $f^{(se)}_n$ is the CPU cycle number needed in the symmetric encryption and HE operation of the symmetric key at the client node $n$. $f_n^{(c)}$ is used to represent the available computing capacity on the $n$-th client node. Thus, the encryption delay can be calculated as
\begin{talign}
    T^{(enc)}_n = \frac{f^{(se)}_n}{f_n^{(c)}}.
\end{talign}

\subsubsection{Encryption Energy} Assume that $\kappa_n^{(c)}$ denotes the effective switched capacitance of $n$-th client node, which can represent the computation energy efficiency of the $n$-th client node. Therefore, the energy consumption of the $n$-th client node for encryption tasks is 
\begin{talign}
    E^{(enc)}_n = \kappa^{(c)}_n f^{(se)}_n (f_n^{(c)})^2.
\end{talign}

\subsubsection{Minimum Security Level} 
We assess FHE robustness using the minimal security level, measured in bits, representing the effort required to breach cryptographic protection. This metric considers three attack vectors: the unique shortest vector problem \cite{peikert2009public}, bounded distance decoding \cite{lyubashevsky2009bounded}, and hybrid dual attack \cite{bi2022hybrid}. The overall privacy for an FHE configuration $(\bm{\lambda}, \bm{q})$ is the minimum security level across these attacks, evaluated via the lattice with error (LWE) \hbox{estimator \cite{cousins2018implementing}}.

Quantifying the relationship between FHE parameters and privacy protection is challenging due to the intricate computational models underlying the LWE Estimator. Parameters like polynomial degree $\lambda_n$ and coefficient modulus $q_n$ jointly influence the cryptographic strength. Their effects are non-linear and context-dependent, making direct analysis complex. 

Recall that we have fixed the modulus $q_n$ across client nodes for simplicity. Therefore, we can focus on the impact of $\lambda_n$ on the minimum security level. The relationship between the polynomial degree $\lambda_n$ and the minimum security level of the $n$-th client node is modeled using a function $f^{(msl)} (\lambda_n)$. The specific expression of the function $f^{(msl)} (\lambda_n)$ will be given in the numerical results section. Here, the function $f^{(msl)} (\lambda_n)$ reflects the security level determined by the LWE Estimator for a given $\lambda_n$, capturing how adjustments to $\lambda_n$ impact the cryptographic strength of the FHE scheme.

Different client nodes within a network often handle varying levels of sensitive data, leading to diverse security level requirements. To accommodate this heterogeneity, we introduce a weight parameter $\varsigma_n$ for each client node $n$. This parameter represents the importance of privacy for that specific device, where higher values of $\varsigma_n$ indicate a greater need for protection. The overall minimum security level of the system is then calculated as the weighted sum of the individual client privacy levels:
\begin{talign}
    U_{msl} = \sum_{n=1}^N \varsigma_n f^{(msl)} (\lambda_n).
\end{talign}

\subsection{Uplink Transmission Phase Analysis}
In this section, the uplink wireless transmission cost is discussed. Once the $n$-th client node finishes the encryption tasks, it transmits the encrypted data to the selected server node. For the sake of simplicity, we assume the $n$-th client node only connects to the server node. The uplink transmission is done via wireless communications, and frequency division multiple access (FDMA) \cite{myung2006single} is used. $B_{total}$ is used to denote the total available bandwidth of the server node. Based on the Shannon formula \cite{shannon1948mathematical}, the uplink transmission rate $r_n$ between the $n$-th client node and the server node is given as follows:
\begin{talign}
    r_n = b_n \log_2 (1 + \frac{p_n g_n}{N_0 b_n}),
\end{talign}
where $b_n$ is the bandwidth between the $n$-th client node and the server node, $p_n$ is the transmit power of the $n$-th client node, $g_n$ is the channel attenuation between the $n$-th client node and the server node, $N_0$ is the noise power spectral density.

\subsubsection{Transmission Delay} Given the transmission rate $r_n$ of the $n$-th client node, the transmission delay can be calculated as
\begin{talign}
    T^{(tr)}_n = \frac{d^{(tr)}_n}{r_n} = \frac{d^{(tr)}_n}{b_n \log_2 (1 + \frac{p_n g_n}{N_0 b_n})},
\end{talign}
where $d^{(tr)}_n$ is the transmitted encrypted data bits from the $n$-th client node.

\subsubsection{Transmission Energy} Once the transmission delay $T^{(tr)}_n$ is calculated, the energy can be written as the product of the transmission power $p_n$ of the $n$-th client node and the transmission delay $T^{(tr)}_n$, which is given as
\begin{talign}
    E^{(tr)}_n = p_n T^{(tr)}_n = \frac{p_n d^{(tr)}_n}{r_n}.
\end{talign}

\subsection{Server Computation Phase Analysis} In this part, we discuss how to formulate the costs during the server computation phase. FHE is considered in the server computation phase. It is well known that the computations on ciphertext consume more computing resources than computations on plaintext. Therefore, it's important to optimize the computation resources during the server computation phase to achieve more efficient performance. However, the computation consumption in this phase is generally hard to determine exactly. Thanks to the estimation function proposed in \cite{li2024privtuner}, we can estimate the computation delay and energy consumption by counting the needed CPU cycle number during computations on the ciphertext. 
Since the estimation function is obtained by performing practical server computation tasks in \cite{li2024privtuner}, for consistency, we assume that our serve computation tasks (i.e., encrypted prediction) are the same as those in \cite{li2024privtuner}. 

\subsubsection{Computation Delay} We define $f^{(cmp)}(\lambda_n)$ and $f^{(eval)}(\lambda_n)$ to denote the total needed CPU cycles per sample for server computation tasks where various operations are involved and server transciphering operation, respectively. $f^{(s)}_n$ is used to represent the allocated computation resource to the $n$-th client node by the server. Thus, the computation delay for the $n$-th client node's tasks during the server computation phase is 
\begin{talign}
    T^{(cmp)}_n = \frac{(f^{(cmp)}(\lambda_n)+f^{(eval)}(\lambda_n)) d^{(cmp)}_n}{\varrho_n f^{(s)}_n},
\end{talign}
where $\varrho_n$ is the number of tokens per sample, and $d^{(cmp)}_n$ is the number of tokens from the client node $n$.

\subsubsection{Computation Energy} Based on the above analysis, the computation energy consumption for the $n$-th client node's tasks during the server computation phase is given as
\begin{talign}
    E^{(cmp)}_n = \frac{\kappa^{(s)} (f^{(cmp)}(\lambda_n)+f^{(eval)}(\lambda_n)) d^{(cmp)}_n (f^{(s)}_n)^2}{\varrho_n},
\end{talign}
where $\kappa^{(s)}$ is the effective switched capacitance of the server.

\subsection{Total Cost Analysis}
The total system delay is the maximum delay experienced by one single client node, encompassing the time required for client-side encryption, wireless transmission to the server, and server-side computation for its tasks. Therefore, the system delay is given as
\begin{talign}
    T_{total} = \max \left\{T^{(enc)}_n + T^{(tr)}_n + T^{(cmp)}_n\right\}.
\end{talign}
The total system energy consumption is the summation of all energy consumption of client nodes and the server. Its expression is shown as follows:
\begin{talign}
    E_{total} = \sum_{n=1}^N \left(E^{(enc)}_n + E^{(tr)}_n + E^{(cmp)}_n\right).
\end{talign}

\section{Studied Optimization Problem Formulation}\label{sec_optiproblem}
In this study, we want to maximize the QKD network utility and minimum security level, and minimize the system costs, including delay and energy consumption from the client encryption phase to the server computation phase. The optimization variables are $\bm{\phi}, \bm{w}, \bm{\lambda}, \bm{p}, \bm{b}, \bm{f}^{(c)}, \bm{f}^{(s)}$. Before formulating the optimization problem, we note that there is one ``maximize'' operation in $T_{total}$. By adding an auxiliary variable $T$, we can limit it to no less than the summation of $T^{(enc)}_n + T^{(tr)}_n + T^{(cmp)}_n$. The studied optimization problem is
\begin{subequations}\label{prob1}
\begin{align}
\mathbb{P}_{1}:&\max\limits_{\bm{\phi},\bm{w},\bm{\lambda},\bm{p},\bm{b},\bm{f}^{(c)},\bm{f}^{(s)}, T}  \alpha_{qkd}U_{qkd} + \alpha_{msl}U_{msl} - \alpha_{t}T \nonumber\\
&\quad \quad \quad \quad \quad \quad \quad \quad - \alpha_{e}E_{total}
\tag{\ref{prob1}}\\
\text{s.t.} \quad 
& \phi_n \geq \phi^{(min)}_n, ~\forall n \in \mathcal{N},\label{cons_phi} \\
& w_l \in (0,1], ~\forall l \in \mathcal{L},\label{cons_w} \\
& \sum_{n=1}^N a_{ln} \phi_n \leq \beta_l (1 - w_l), \forall l \in \mathcal{L}. \label{cons_linkcapacity} \\
& \lambda_n \in \left\{\lambda^{(set)}_1, \lambda^{(set)}_2, \cdots, \lambda^{(set)}_M\right\}, ~\forall n \in \mathcal{N}, \label{cons_lambda} \\
& p_n \leq p^{(max)}_n, ~\forall n \in \mathcal{N}, \label{cons_p} \\
& \sum_{n=1}^N b_n \leq B_{total}, \label{cons_b} \\
& f^{(c)}_n \leq f^{(max)}_n, ~\forall n \in \mathcal{N}, \label{cons_fc} \\
& \sum_{n=1}^N f^{(s)}_n \leq f_{total}, \label{cons_fs} \\
& T^{(enc)}_n + T^{(tr)}_n + T^{(cmp)}_n \leq T, ~\forall n \in \mathcal{N}. \label{cons_T}
\end{align}
\end{subequations}

\subsection{Parameter and Constraint Illustration}
In Problem $\mathbb{P}_{1}$ of (\ref{prob1}) above, $\alpha_{qkd}$, $\alpha_{msl}$, $\alpha_{t}$, and $\alpha_{e}$ are the weight parameters of $U_{qkd}$, $U_{msl}$, $T_{total}$, and $E_{total}$, respectively. Those weight parameters are used to adjust the metrics' value scale, which is helpful for optimization effectiveness. Constraint (\ref{cons_phi}) means that the allocated quantum entanglement rate of the $n$-th client node should meet the minimum rate requirement of this node, and $\phi^{(min)}_n$ is the minimum rate needed for the $n$-th client node. Constraint (\ref{cons_w}) is the fidelity bounds of the Werner parameter $w_l$. Constraint (\ref{cons_linkcapacity}) means that the total allocated entanglement rate can't be greater than the maximum entanglement generation rate of one link. Constraint (\ref{cons_lambda}) is the discrete value set of $\lambda_n$ and $\lambda^{(set)}_1\leq\lambda^{(set)}_2\leq \cdots\leq \lambda^{(set)}_M$. Constraint (\ref{cons_p}) limits the transmit power at the client node. Constraints (\ref{cons_b}) and (\ref{cons_fs}) mean that the summation of the allocated bandwidth and computation resources for each client node by the server can't exceed the total available bandwidth and computation resources. Constraint (\ref{cons_fc}) limits the computation resource at the client node. Constraint (\ref{cons_T}) limits the system delay.

\subsection{Non-Convexity and NP-Hardness}
There are many coupled product and ratio terms in the objective function and constraints, which are commonly considered as multiplicative programming or fractional programming. Besides, $\bm{\lambda}$ is a discrete variable, leading the optimization problem to be a mixed-integer on-linear programming (MINLP). Given those terms and variables, Problem (\ref{prob1}) is non-convex and NP-hard.

\section{Proposed QuHE Algorithm to Solve the Optimization Problem}\label{sec_quhe}
In this section, the proposed QuHE algorithm is presented to solve the Problem (\ref{prob1}). We consider using three-stage alternating optimization to tackle this difficult optimization. Assume that we are in the $(i+1)$-th iteration, and the algorithm procedure is given as follows:
\begin{itemize}
    \item Stage 1: Fix $\bm{\lambda}^{(i)},\bm{p}^{(i)},\bm{b}^{(i)},(\bm{f}^{(c)})^{(i)},(\bm{f}^{(s)})^{(i)}, T^{(i)}$, and then optimize $\bm{\phi}^{(i+1)}$ and $\bm{w}^{(i+1)}$.
    \item Stage 2: Fix $\bm{\phi}^{(i+1)},\bm{w}^{(i+1)},\bm{p}^{(i)},\bm{b}^{(i)},(\bm{f}^{(c)})^{(i)},(\bm{f}^{(s)})^{(i)}$, and then optimize $\bm{\lambda}^{(i+1)}, T^{(i)}_{s_2}$. Note that $T^{(i)}_{s_2}$ is not the final value in the $(i+1)$-th iteration, and it will be optimized in Stage 3.
    \item Stage 3: Fix $\bm{\phi}^{(i+1)},\bm{w}^{(i+1)},\bm{\lambda}^{(i+1)}$, and then optimize $\bm{p}^{(i+1)},\bm{b}^{(i+1)},(\bm{f}^{(c)})^{(i+1)},(\bm{f}^{(s)})^{(i+1)}, T^{(i+1)}$.
\end{itemize}
Repeat those steps, and the QuHE algorithm will converge under certain accuracy conditions. Next, we illustrate the details at each stage.

\subsection{Stage 1 of the Proposed QuHE Algorithm}
If given $\bm{\lambda},\bm{p},\bm{b},\bm{f}^{(c)},\bm{f}^{(s)}, T$, the remaining optimization variables are $\bm{\phi}$ and $\bm{w}$. Note that the objective function in Problem (\ref{prob1}) increases monotonically with $\varpi_n$. Recall that $\varpi_n = \prod_{l=1}^L w_l^{a_{ln}}$, and we know that the objective function in Problem (\ref{prob1}) also increases monotonically with $\bm{w}$. Therefore, the optimal $\bm{w}$ is the maximum value that $\bm{w}$ can take. From Constraint (\ref{cons_linkcapacity}), we get the optimal value of $w_l^\star$ is 
\begin{talign}
    w_l^\star = 1 - \frac{\sum_{n=1}^N a_{ln} \phi_n}{\beta_l}, \label{eq_w_optimal}
\end{talign}
which also satisfies Constraint (\ref{cons_w}). However, the term $U_{qkd}$ is still a multiplicative term, which is hard to analyze. Thus, we perform the logarithmic operation on the objective function to transform the multiplicative term into the summation term, which is easier to analyze. Since the term $(\alpha_{msl}U_{msl} - \alpha_{t}T - \alpha_{e}E_{total})$ is a constant at Stage 1 and it is also troublesome to pay attention to its positive condition while using the logarithmic operation, we decide to omit this term in the objective function. Besides, we study the ``minimization'' of Problem (\ref{prob1}). The new optimization problem would be
\begin{subequations}\label{prob2}
\begin{talign}
\mathbb{P}_{2}: \min\limits_{\bm{\phi}}  
& -\sum_{n=1}^N \ln \left( F_{skf}\left(\prod_{l=1}^L {\left(1 - \frac{\sum_{n=1}^N a_{ln} \phi_n}{\beta_l}\right)}^{a_{ln}}\right) \right) \nonumber \\
& -\ln \alpha_{qkd} - \sum_{n=1}^N \ln \phi_n  
\tag{\ref{prob2}}\\
\text{s.t.} \quad & \text{(\ref{cons_phi})}, \nonumber \\
& 0 < \frac{\sum_{n=1}^N a_{ln} \phi_n}{\beta_l} < 1, \forall l \in \mathcal{L}, \label{cons_phi2} \\
& 0.779944 < \prod_{l=1}^L {\left(1 - \frac{\sum_{n=1}^N a_{ln} \phi_n}{\beta_l}\right)}^{a_{ln}}, \forall n \in \mathcal{N}, \label{cons_positiveln}
\end{talign}
\end{subequations}
where Constraint (\ref{cons_phi2}) is Constraint (\ref{cons_w}) when we replace $w_l$ by $w_l^\star$ in it. Constraint (\ref{cons_positiveln}) is introduced to keep the logarithmic function in the objective function positive. It's easy to know that $F_{skf}(w)$ is monotonically increasing over the part of the function value greater than zero. 0.779944 is the largest number that makes $F_{skf}(w) = 0$, which can be obtained by using the graphing calculator Desmos \cite{liang2016teaching}. Furthermore, we introduce a new auxiliary variable $\varphi_n:=\ln(\phi_n), n \in \mathcal{N}$. Define $\bm{\varphi}:=[\varphi_n]|_{n \in \mathcal{N}}$. Therefore, Problem (\ref{prob2}) can be transformed into Problem (\ref{prob3}):
\begin{subequations}\label{prob3}
\begin{talign}
\mathbb{P}_{3}: \min\limits_{\bm{\varphi}}  
& -\sum_{n=1}^N \ln \left( F_{skf}\left(\prod_{l=1}^L {\left(1 - \frac{\sum_{n=1}^N a_{ln} \phi_n}{\beta_l}\right)}^{a_{ln}}\right) \right) \nonumber \\
& -\ln \alpha_{qkd} - \sum_{n=1}^N \ln \phi_n  
\tag{\ref{prob3}}\\
\text{s.t.} \quad & e^{\varphi_n} > \phi_n^{(min)}, \forall n \in \mathcal{N},\label{cons_varphi} \\
& 0 < \frac{\sum_{n=1}^N a_{ln} e^{\varphi_n}}{\beta_l} < 1, \forall l \in \mathcal{L}, \label{cons_varphi2} \\
& 0.779944 < \prod_{l=1}^L {\left(1 - \frac{\sum_{n=1}^N a_{ln} e^{\varphi_n}}{\beta_l}\right)}^{a_{ln}}, \forall n \in \mathcal{N}. \label{cons_positiveln_varphi}
\end{talign}
\end{subequations}
Based on Proposition 1, Theorem 1, and Theorem 2 in \cite{kar2024convexification}, it's known that Problem (\ref{prob3}) is convex, which common convex tools can solve. Therefore, we obtain the optimal solution $\bm{\varphi}^\star$, and we further get $\bm{\phi}^\star = e^{\bm{\varphi}^\star}$ and $\bm{w}^\star$ by Equation (\ref{eq_w_optimal}). The procedure of Stage 1 in the proposed QuHE algorithm is shown in Algorithm \ref{algo:QuHE_stage1}.
\begin{algorithm}
\caption{Stage 1 of the Proposed QuHE Algorithm.}
\label{algo:QuHE_stage1}
Initialize a feasible point $\bm{\phi}^{(0)}$;

Obtain the optimal solution $\bm{\varphi}^\star$ by solving Problem (\ref{prob3}) via common convex tools;

Let $\bm{\phi}^\star = e^{\bm{\varphi}^\star}$;

Obtain $\bm{w}^\star$ by the equation (\ref{eq_w_optimal});
\end{algorithm}

\subsection{Stage 2 of the Proposed QuHE Algorithm}
Once $\bm{\phi},\bm{w},\bm{p},\bm{b},\bm{f}^{(c)},\bm{f}^{(s)}$ are fixed, the optimization variables are $\bm{\lambda}$ and $T$. The objective function in Problem (\ref{prob1}) decreases monotonically with $T$. From Constraint (\ref{cons_T}), we obtain the optimal value $T_{s_2}$ with given $\lambda_n$, whose expression is 
\begin{talign}
    T_{s_2} = \frac{f^{(se)}_n}{f_n^{(c)}} + \frac{d^{(tr)}_n}{r_n} +  \frac{(f^{(cmp)}(\lambda_n)+f^{(eval)}(\lambda_n)) d^{(cmp)}_n}{\varrho_n f^{(s)}_n}.
\end{talign}
If we plug $T_{s_2}$ into the objective function in Problem (\ref{prob1}), the optimization would become only with the variable $\bm{\lambda}$. The original optimization problem (\ref{prob1}) is simplified as
\begin{subequations}\label{prob4}
\begin{align}
\mathbb{P}_{4}:&\max\limits_{\bm{\lambda}}  ~~ \alpha_{qkd}U_{qkd} + \alpha_{msl}U_{msl} - \alpha_{t}T_{s_2} - \alpha_{e}E_{total}
\tag{\ref{prob4}}\\
\text{s.t.} 
& \quad  \text{(\ref{cons_lambda})}.\nonumber
\end{align}
\end{subequations}
Recall that $\bm{\lambda}$ is a discrete variable, and we can use a simple exhaustive search method to find the optimal value $\bm{\lambda}^\star$. However, the complexity of the exhaustive search method will increase significantly with the increase of the search space. Therefore, we try to use the branch and bound technique to find the optimal value $\bm{\lambda}^\star$. Since the branch and bound is a mature technique, we don't give too much illustration here, and interested readers can refer to \cite{lawler1966branch} for further information. 

For brevity, define the objective function in Problem (\ref{prob4}) as $F_{s_2}(\bm{\lambda})$. When we get the optimal $\bm{\lambda}$, i.e., $\bm{\lambda}^\star$, we can also obtain the optimal value of $T$ at the stage 2, which is given as
\begin{talign}
    T_{s_2}^\star = \frac{f^{(se)}_n }{f_n^{(c)}} + \frac{d^{(tr)}_n}{r_n} +  \frac{(f^{(cmp)}(\lambda_n^\star)+f^{(eval)}(\lambda_n^\star)) d^{(cmp)}_n}{\varrho_n f^{(s)}_n}.\label{eq_T_s2}
\end{talign}
The detailed Stage 2 algorithm procedure is presented in Algorithm \ref{algo:QuHE_stage2}.
\begin{algorithm}
\caption{Stage 2 of the Proposed QuHE Algorithm.}
\label{algo:QuHE_stage2}
Initialize the priority queue \(Q\) with the initial search space as an empty partial solution \(\bm{\lambda}_{\text{partial}} = \emptyset\), with an initial upper bound of \(+\infty\);

Set the initial best solution \(\bm{\lambda}^\star = \bm{0}\) and objective value \(F_{s_2}^\star = -\infty\);

Extract the subproblem with the highest upper bound from \(Q\);

Represent the subproblem as a partial solution \(\bm{\lambda}_{\text{partial}}\);

If \(\bm{\lambda}_{\text{partial}}\) has all variables assigned:
    Compute \(F_{s_2}(\bm{\lambda}_{\text{partial}})\);
    If \(F_{s_2}(\bm{\lambda}_{\text{partial}}) > F_{s_2}^\star\), update \(F_{s_2}^\star \gets F_{s_2}(\bm{\lambda}_{\text{partial}})\) and \(\bm{\lambda}^\star \gets \bm{\lambda}_{\text{partial}}\);

Otherwise, perform branching:
    Select the next variable \(\lambda_n\) to assign;
    For each value \(v \in \{\lambda_1^{\text{set}}, \ldots, \lambda_M^{\text{set}}\}\):
        Create a new partial solution by setting \(\lambda_n = v\);
        Compute an upper bound for the new subproblem;
        If the upper bound \(> F_{s_2}^\star\), add the new subproblem to \(Q\);
        Otherwise, prune the subproblem;

Repeat the process until \(Q\) is empty;

Obtain optimal solution \(\bm{\lambda}^\star\) and objective value \(F_{s_2}^\star\).

Obtain optimal solution $T_{s_2}^\star$ via Equation (\ref{eq_T_s2}).
\end{algorithm}

\subsection{Stage 3 of the Proposed QuHE Algorithm}
In Stage 3, we fix $\bm{\phi}, \bm{w}, \bm{\lambda}$, and then optimize $\bm{p}, \bm{b}, \bm{f}^{(c)}, \bm{f}^{(s)}, T$. Since $\alpha_{qkd} U_{qkd}$ and $\alpha_{msl} U_{msl}$ are constant terms in Stage 3, we can rewrite the optimization problem (\ref{prob1}) as follows:
\begin{subequations}\label{prob5}
\begin{align}
\mathbb{P}_{5}:&\max\limits_{\bm{p},\bm{b},\bm{f}^{(c)},\bm{f}^{(s)}, T}  - \alpha_{e} \sum_{n=1}^N \left(\kappa^{(c)}_n f^{(se)}_n(f_n^{(c)})^2\right) \nonumber \\
&\!\!\!\!\!\!\!\!\!\!\!\!\!\! - \alpha_{e} \sum_{n=1}^N \left(\frac{\kappa^{(s)} (f^{(cmp)}(\lambda_n)+f^{(eval)}(\lambda_n)) d^{(cmp)}_n (f^{(s)}_n)^2}{\varrho_n}\right) \nonumber \\
& \!\!\!\!\!\!\!\!\!\!\!\!\!\!- \alpha_{e} \sum_{n=1}^N \left(\frac{p_n d^{(tr)}_n}{r_n}\right) - \alpha_{t}T 
\tag{\ref{prob5}}\\
\text{s.t.} \quad
& \text{(\ref{cons_p})}, \text{(\ref{cons_b})}, \text{(\ref{cons_fc})}, \text{(\ref{cons_fs})}, \text{(\ref{cons_T})}.\nonumber
\end{align}
\end{subequations}
It's known that $r_n$ is jointly concave to $b_n$ and $p_n$ \cite{qian2024user}. Therefore, the term $d^{(tr)}_n/r_n$ in Constraint (\ref{cons_T}) is convex, which is based on the chain role in \cite{boyd2004convex}. We further know that Constraint (\ref{cons_T}) is convex. It's easy to get that other constraints are all convex, and the only non-concave term in the objective function in Problem (\ref{prob5}) is $- \alpha_e \sum_{n=1}^N \frac{p_n d^{(tr)}_n}{r_n}$. Next, we present how to make this term concave. Let 
\begin{talign}
    z_n=\frac{1}{2 p_n d_n^{(tr)} r_n}, \label{eq_zn}
\end{talign}
and $\bm{z}:=[z_n]|_{n \in \mathcal{N}}$, and we do the following transformation:
\begin{talign}
    \frac{p_n d_n^{(tr)}}{r_n} \rightarrow \left(p_n d_n^{(tr)}\right)^2 z_n + \frac{1}{4 r_n^2 z_n},\label{eq_fp}
\end{talign}
where the right side is proved to be convex to $p_n$ and $b_n$ with fixed $z_n$ (refer to Section \uppercase\expandafter{\romannumeral4} in \cite{10368052}). Besides, since the term $\frac{p_n d^{(tr)}_n}{r_n}$ is pseudoconvex to $p_n$ and $b_n$ \cite{shen2018fractional}, we can obtain optimal solutions of $p_n$ and $b_n$ by alternatively optimize $z_n$ and $(p_n, b_n)$ \cite{10368052}. We define the following function:
\begin{talign}
    f_n^{(tr)}(b_n, p_n, z_n) = \left(p_n d_n^{(tr)}\right)^2 z_n + \frac{1}{4 r_n^2 z_n}.
\end{talign}
Problem (\ref{prob5}) can be rewritten as
\begin{subequations}\label{prob6}
\begin{align}
\mathbb{P}_{6}:&\max\limits_{\bm{p},\bm{b},\bm{f}^{(c)},\bm{f}^{(s)}, T, \bm{z}}  - \alpha_{e} \sum_{n=1}^N \left(\kappa^{(c)}_n f^{(se)}_n (f_n^{(c)})^2\right) \nonumber \\
&\!\!\!\!\!\!\!\!\!\!\!\!\!\! - \alpha_{e} \sum_{n=1}^N \left(\frac{\kappa^{(s)} (f^{(cmp)}(\lambda_n)+f^{(eval)}(\lambda_n)) d^{(cmp)}_n (f^{(s)}_n)^2}{\varrho_n}\right) \nonumber \\
&\!\!\!\!\!\!\!\!\!\!\!\!\!\! - \alpha_{e} \sum_{n=1}^N f_n^{(tr)}(b_n, p_n, z_n) - \alpha_{t}T 
\tag{\ref{prob6}}\\
\text{s.t.} \quad
& \text{(\ref{cons_p})}, \text{(\ref{cons_b})}, \text{(\ref{cons_fc})}, \text{(\ref{cons_fs})}, \text{(\ref{cons_T})}.\nonumber
\end{align}
\end{subequations}
Now, the objective function in Problem (\ref{prob6}) is concave if we fix $\bm{z}$, and if we fix $\bm{p}$, $\bm{b}$, $\bm{f}^{(c)}$, $\bm{f}^{(s)}$, it is also concave to $\bm{z}$. Therefore, we can alternatively optimize $(\bm{z})$ and $\bm{p}$, $\bm{b}$, $\bm{f}^{(c)}$, $\bm{f}^{(s)}$, which can be solved by common convex tools. The procedure in Stage 3 is shown in Algorithm \ref{algo:QuHE_stage3}.
\begin{algorithm}
\caption{Stage 3 of the Proposed QuHE Algorithm.}
\label{algo:QuHE_stage3}

Initialize $i \leftarrow -1$ and a feasible point $(\bm{b}^{(0)},\bm{p}^{(0)},(\bm{f}^{(c)})^{(0)},(\bm{f}^{(s)})^{(0)})$;

\Repeat{Function value in optimization (\ref{prob6}) convergences}{
Let $i \leftarrow i+1$;

Update $\bm{z}^{(i+1)}$ with $(\bm{b}^{(i)},\bm{p}^{(i)})$ by Equation (\ref{eq_zn});

Update $(\bm{b}^{(i+1)},\bm{p}^{(i+1)},(\bm{f}^{(c)})^{(i+1)},(\bm{f}^{(s)})^{(i+1)}, T^{(i+1)})$ by solving Problem (\ref{prob6}) with fixed $\bm{z}^{(i+1)}$;
}
Obtain optimal solutions $\bm{b}^\star,\bm{p}^\star,(\bm{f}^{(c)})^\star,(\bm{f}^{(s)})^\star, T^\star$.
\end{algorithm}

\subsection{Whole Procedure of the Proposed QuHE Algorithm}
The proposed QuHE algorithm consists of alternative optimization stages in three blocks, i.e., $(\bm{\phi}, \bm{w})$, $(\bm{\lambda}, T)$, $(\bm{p}, \bm{b}, \bm{f}^{(c)}, \bm{f}^{(s)}, T)$. Note that $T$ is updated twice at Stage 2 and Stage 3 because $T$ is associated with optimization variables at those two stages. We give the whole procedure of the proposed QuHE algorithm in Algorithm \ref{algo:QuHE}.
\begin{algorithm}
\caption{The Whole Procedure of the Proposed QuHE Algorithm.}
\label{algo:QuHE}

Initialize $i \leftarrow -1$ and a feasible point $\left(\bm{\phi}^{(0)},\bm{w}^{(0)},\bm{\lambda}^{(0)},\bm{b}^{(0)},\bm{p}^{(0)},(\bm{f}^{(c)})^{(0)},(\bm{f}^{(s)})^{(0)},T^{(0)}\right)$;

\Repeat{Function value in optimization (\ref{prob1}) convergences}{
Let $i \leftarrow i+1$;

\textbf{Stage 1}: Fix $\bm{\lambda}^{(i)}$, $\bm{b}^{(i)}$, $\bm{p}^{(i)}$, $(\bm{f}^{(c)})^{(i)}$, $(\bm{f}^{(s)})^{(i)}$, $T^{(i)}$, and obtain $\bm{\phi}^{(i+1)}$, $\bm{w}^{(i+1)}$ by using Algorithm \ref{algo:QuHE_stage1};

\textbf{Stage 2}: Fix $\bm{\phi}^{(i+1)}$, $\bm{w}^{(i+1)}$, $\bm{b}^{(i)}$, $\bm{p}^{(i)}$, $(\bm{f}^{(c)})^{(i)}$, $(\bm{f}^{(s)})^{(i)}$, and obtain $\bm{\lambda}^{(i+1)}$ and $T^{(i+1)}_{s_2}$ by using Algorithm \ref{algo:QuHE_stage2}; 

\textbf{Stage 3}: Fix $\bm{\phi}^{(i+1)}$, $\bm{w}^{(i+1)}$, $\bm{\lambda}^{(i+1)}$, and obtain $\bm{b}^{(i+1)}$, $\bm{p}^{(i+1)}$, $(\bm{f}^{(c)})^{(i+1)}$, $(\bm{f}^{(s)})^{(i+1)}$, $T^{(i+1)}$ by running Algorithm \ref{algo:QuHE_stage3};
}
Obtain optimal solutions $\bm{\phi}^\star$, $\bm{w}^\star$, $\bm{\lambda}^\star$, $\bm{b}^\star$, $\bm{p}^\star$, $(\bm{f}^{(c)})^\star$, $(\bm{f}^{(s)})^\star$, $T^\star$.
\end{algorithm}

\subsection{Solution Optimality Analysis}
We present the solution optimality analysis of the proposed QuHE algorithm. In Stage 1, the related transformations don't affect the optimality of solutions, i.e., the solutions $\bm{\phi}$ and $\bm{w}$ are optimal in this stage. In Stage 2, since we use the branch and bound technique, which is a common optimization method to find the globally optimal solution, the solution $\bm{\lambda}$ is also optimal. In Stage 3, we only perform the fractional programming transformation of the term $\frac{p_n d^{(tr)}_n}{r_n}$, i.e., Equation (\ref{eq_fp}). A stationary point solution is guaranteed if using this fractional programming technique \cite{10368052}. Besides, we know that the term $\frac{p_n d^{(tr)}_n}{r_n}$ is pseudoconvex to $p_n$ and $b_n$ \cite{shen2018fractional}, and a stationary point solution of a pseudoconvex problem is also the globally optimal solution. Therefore, we can find the globally optimal solutions of $\bm{p}$, $\bm{b}$, $\bm{f}^{(c)}$, $\bm{f}^{(s)}$ by using the fractional programming technique in \cite{10368052}.

Since we always find the optimal solutions in every stage, the solution optimality of the proposed QuHE algorithm is at least a stationary point \hbox{solution \cite{chen2012maximum}}.

\subsection{Complexity Analysis}
In this section, we analyze the complexity of the proposed QuHE algorithm. Assume the solution accuracy tolerance is $\epsilon$ ($\epsilon>0$). In Optimization (\ref{prob1}), there are $N$ variables and $2N+L$ constraints, and the worst-case complexity of solving it is $\mathcal{O}(N^{3.5}+L^{3.5})\log(1/\epsilon)$. There are also $N+L$ equality computations to obtain solutions of $\bm{\phi}$ and $\bm{w}$. In Optimization (\ref{prob4}), there are $N$ variables, and each variable has $M$ discrete value choices. Thus, the worst-case complexity of solving Optimization (\ref{prob4}) by the branch and bound method is $\mathcal{O}(M^N)$ \cite{lawler1966branch}. In Optimization (\ref{prob6}), there are $4N$ variables and $3N+2$ constraints, and the worst-case complexity of solving it is $\mathcal{O}(N^{3.5})\log(1/\epsilon)$. Assume that there are $\mathcal{I}$ iteration needed in the QuHE algorithm. Therefore, the overall worst-case complexity of the proposed QuHE algorithm is $\mathcal{I} \left(\mathcal{O}(N^{3.5}+L^{3.5})\log(1/\epsilon)+\mathcal{O}(M^N)\right)$.
\begin{figure}[tbp]
\centering
\includegraphics[width=0.4\textwidth]{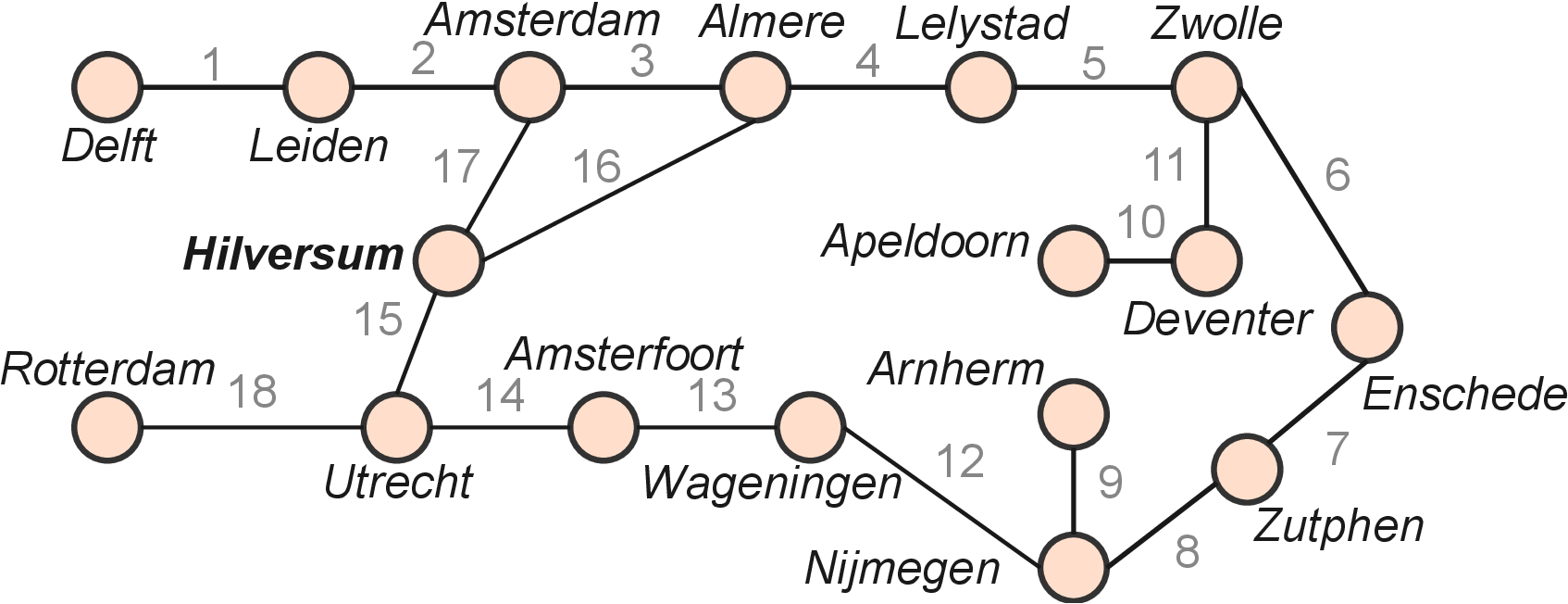}
\vspace{-6pt}\caption{Studied quantum network topology from \cite{vardoyan2024bipartite}.}
\label{fig_topology}
\end{figure}
\begin{table}[t]
    \centering
    \scriptsize 
    \caption{Routes with end nodes and links.}
    \begin{tabular}{|c|c|c|}
        \hline
        Route ID & End nodes  & Links  \\ \hline
        1       & (\textit{Hilversum}, \textit{Delft})    & (17, 2, 1) \\ 
        2       & (\textit{Hilversum}, \textit{Zwolle})    & (17, 3, 4, 5) \\ 
        3       & (\textit{Hilversum}s, \textit{Apeldoorn})   & (16, 4, 5, 11, 10) \\
        4       & (\textit{Hilversum}, \textit{Rotterdam})   & (15, 18) \\
        5       & (\textit{Hilversum}, \textit{Arnherm})   & (15, 14, 13, 12, 9) \\
        6       & (\textit{Hilversum}, \textit{Enschede})    & (15, 14, 13, 12, 8, 7) \\ \hline
    \end{tabular}
    \label{tab:routes}
\end{table}
\begin{table}[t]
    \centering
    \scriptsize 
    \caption{Link lengths and $\beta_j$ for various links.}
    \begin{tabular}{|c|c|c|c|c|c|}
        \hline
        Link ID   & Length (km) & $\beta_j$ & Link ID   & Length (km) & $\beta_j$  \\ \hline
        1                 & 30.6                 & 89.84    &  10                & 24.4                 & 100.98              \\ 
        2                 & 60.4                 & 53.79           &  11                & 44.7                 & 68.75      \\ 
        3                 & 38.9                 & 77.47       &12                & 66.3                 & 49.35             \\ 
        4                 & 44.2                 & 69.44          &13                & 62.5                 & 52.40         \\ 
        5                 & 47.7                 & 65.12          &14                & 33.8                 & 84.63       \\ 
        6                 & 78.7                 & 40.76        &15                & 36.7                 & 80.54           \\ 
        7                 & 60.0                 & 54.17        &16                & 35.4                 & 82.41         \\ 
        8                 & 58.1                 & 56.25         &17                & 30.2                 & 90.52        \\ 
        9                 & 25.7                 & 99.02          &18                & 70.0                 & 46.82                   \\ \hline
    \end{tabular}
    \label{tab:links}
\end{table}
\section{Numerical Results} \label{sec_simulation}
In this section, we present the numerical results.

\subsection{Parameter Setting}
We utilize the SURFnet topology \cite{rabbie2022designing}, a real-world backbone fibre network for research, to simulate QKD service. We let $N=6$, and choose six routes with end nodes and links in Table \ref{tab:routes}. Node \textit{Hilversum} is selected as the QKD key center. We give the value of $\beta_j$ and link lengths in \hbox{Table \ref{tab:links}}, and let $L=18$. The minimum rate needed for the $n$-th client node $\phi^{(min)}_n$ is set as $0.5$ pairs per second. The set of $\lambda$ is $\{2^{15}, 2^{16}, 2^{17}\}$. Expressions of functions $f^{(eval)}(\lambda_n)$, $f^{(msl)}(\lambda_n)$, and $f^{(cmp)}(\lambda_n)$ is presented as follows:
\begin{talign}
    &f^{(eval)}(\lambda_n) = 0.012(\lambda_n + 64500)^2,\\
    &f^{(msl)}(\lambda_n) = 0.002\lambda_n + 1.4789,\\
    &f^{(cmp)}(\lambda_n) = 8917959.4\lambda_n - 51292440000.
\end{talign}
These functions are obtained by curve fitting via running the CKKS mechanism and the LWE-estimator under three attacks (i.e., uSVP, BDD, and hybrid dual) in \cite{li2024privtuner}. Set encryption token number $d^{(cmp)}_n$ as 160, transmit data size $d^{(tr)}_n$ as $3\times10^9$ bits, token number per sample $\varrho_n$ as 10, and the CPU cycles of the client's encryption work $f^{(se)}_n$ as $10^6$. The total computation resource at the server side $f_{total}$ is 20 GHz. The total computation resources at the client side $f^{(max)}_n$ are \hbox{3 GHz}. The total available bandwidth $B_{total}$ is 10 MHz. The maximum transmit power at the client node $p^{(max)}_n$ is \hbox{0.2 W}. The effective switched capacitance of the client or server node (i.e., $\kappa_n^{(c)}$ and $\kappa_n^{(s)}$) is $10^{-28}$. We employ the model $128.1+37.6\log_{10} (\text{distance})$ as the large-scale fading between the client node and the server. Rayleigh fading is used as the small-scale fading. The distance between the client node and the server is randomly chosen in a circular network topology with a radius of 1000 meters. The weight parameters $\alpha_{qkd}$, $\alpha_{msl}$, $\alpha_t$, and $\alpha_e$ are set as 1, $10^{-2}$, $10^{-4}$, and $10^{-4}$, respectively. The weight parameters of the privacy importance at the client nodes $\{\varsigma_1, \varsigma_2, \cdots, \varsigma_6\}$ are $\{0.1,0.1,0.1,0.2,0.2,0.3\}$. The solution accuracy tolerance $\epsilon$ is set as $10^{-4}$. Simulations are conducted by Matlab 2021b with CVX tools. The hardware configuration is given as follows: A 3.8 GHZ Intel(R) Xeon(R) W-2235 CPU and 32 GB RAM.

\subsection{Baseline Selection}

For Stage 1, we use gradient descent (learning rate 0.01), simulated annealing (via Matlab’s \textit{simulannealbnd} function), and random selection, which samples $10^4$ points uniformly from the feasible space and selects the best based on Problem (\ref{prob3})’s objective.

For the whole algorithm procedure, we select \underline{a}verage \underline{a}llocation (AA), \underline{o}ptimize $\bm{\lambda}$ only with \underline{a}verage \underline{a}llocation (OLAA), \underline{o}ptimize \underline{c}omputation and \underline{c}ommunication \underline{r}esources only (OCCR) as baselines. In baseline AA, $\lambda_n$ is $2^{15}$, $p_n$ is $p^{(max)}_n$, $b_n$ is set as $B_{total}/N$, $f^{(c)}_n$ is $f^{(max)}_n$, and $f^{(s)}_n$ is $f_{total}/N$. In baseline OLAA, we only optimize $\lambda_n$ using the QuHE algorithm in Stage 2 and average allocate the communication and computation resources. In baseline OCCR, we optimize the communication and computation resources using the QuHE algorithm in Stage 3 and fix $\lambda_n$ as $2^{15}$.
\begin{figure}[t]
\centering
\subfigure[Function values across samples.]{
    \includegraphics[width=0.22\textwidth]{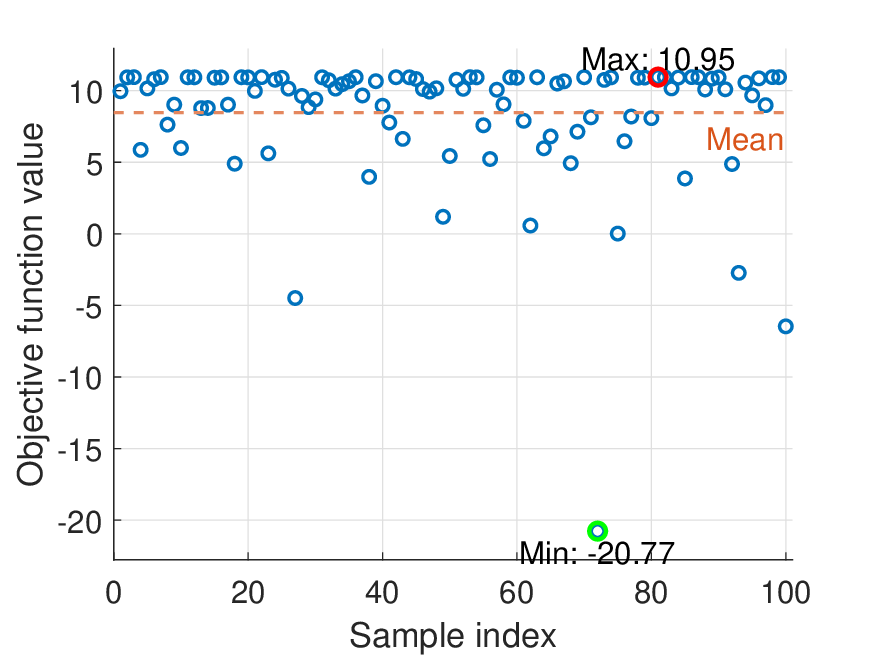}
    \label{fig_optimality_gap_sample_dot}
}
\subfigure[Distribution of the function values.]{
    \includegraphics[width=0.22\textwidth]{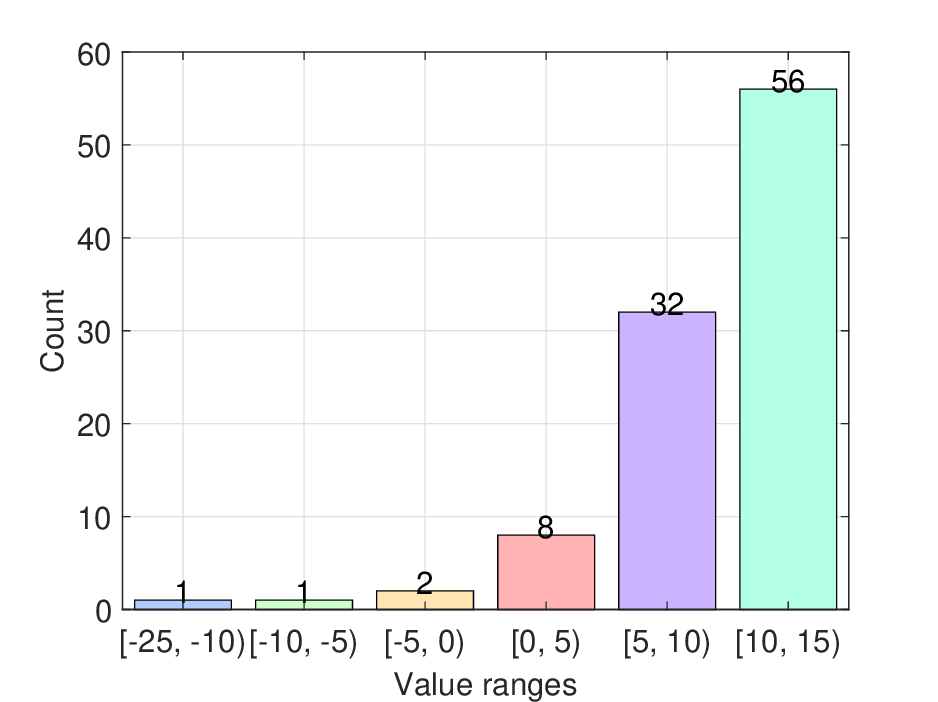}
    \label{fig_optimality_gap_sample_bar}
}
\vspace{-6pt}
\caption{Optimality analysis in 100 samples.}
\label{fig_optimality}
\end{figure}
\begin{table}[t]
\centering
\scriptsize 
\caption{\( \bm{\phi} \) values of different methods.}
\label{tab:phi_comparison}
\begin{tabular}{|c|c|c|c|c|}
\hline
$\phi_n$ & QuHE Stage 1 & Gradient descent & Sim. annealing & Random select \\ \hline
$\phi_1$  & \textbf{2.098}        & \textbf{2.098}            & 2.035          & 1.926         \\ \hline
$\phi_2$  & 1.106        & 1.106            & 1.043          & \textbf{1.442}         \\ \hline
$\phi_3$  & 1.103        & 1.103            & 0.9103         & \textbf{2.045}         \\ \hline
$\phi_4$  & 1.872        & 1.872            & \textbf{1.886}          & 1.442         \\ \hline
$\phi_5$  & 0.6864       & 0.6864           & 0.7975         & \textbf{1.001}         \\ \hline
$\phi_6$  & 0.5781       & 0.5781           & 0.6168         & \textbf{1.151}         \\ \hline
\end{tabular}
\end{table}
\begin{table}[t]
\centering
\scriptsize 
\caption{\( \bm{w} \) values of different methods}
\label{tab:w_comparison}
\resizebox{\columnwidth}{!}{%
\begin{tabular}{|c|c|c|c|c|}
\hline
$w_l$ & QuHE Stage 1 & Gradient descent & Sim. annealing & Random select \\ \hline
$w_1$      & 0.9766       & 0.9766           & 0.9773         & \textbf{0.9786}        \\ \hline
$w_2$      & 0.9610       & 0.9610           & 0.9622         & \textbf{0.9642}        \\ \hline
$w_3$      & 0.9857       & 0.9857           & \textbf{0.9865}         & 0.9814        \\ \hline
$w_4$      & 0.9682       & 0.9682           & \textbf{0.9719}         & 0.9498        \\ \hline
$w_5$      & 0.9661       & 0.9661           & \textbf{0.9700}         & 0.9465        \\ \hline
$w_6$      & \textbf{1.0000}       & \textbf{1.0000}           & \textbf{1.0000}         & \textbf{1.0000}        \\ \hline
$w_7$      & \textbf{0.9893}       & \textbf{0.9893}           & 0.9886         & 0.9787        \\ \hline
$w_8$      & \textbf{0.9897}       & \textbf{0.9897}           & 0.9890         & 0.9795        \\ \hline
$w_9$      & \textbf{0.9931}       & \textbf{0.9931}           & 0.9919         & 0.9899        \\ \hline
$w_{10}$   & 0.9891       & 0.9891           & \textbf{0.9910}         & 0.9797        \\ \hline
$w_{11}$   & 0.9840       & 0.9840           & \textbf{0.9868}         & 0.9703        \\ \hline
$w_{12}$   & \textbf{0.9744}       & \textbf{0.9744}           & 0.9713         & 0.9564        \\ \hline
$w_{13}$   & \textbf{0.9759}       & \textbf{0.9759}           & 0.9730         & 0.9589        \\ \hline
$w_{14}$   & \textbf{0.9851}       & \textbf{0.9851}           & 0.9833         & 0.9746        \\ \hline
$w_{15}$   & \textbf{0.9611}       & \textbf{0.9611}           & 0.9590         & 0.9554        \\ \hline
$w_{16}$   & 0.9866       & 0.9866           & \textbf{0.9890}         & 0.9752        \\ \hline
$w_{17}$   & 0.9646       & 0.9646           & \textbf{0.9660}         & 0.9628        \\ \hline
$w_{18}$   & 0.9600      & 0.9600           & 0.9597         & \textbf{0.9692}        \\ \hline
\end{tabular}
}
\end{table}
\subsection{Optimality Analysis}
To evaluate the robustness and reliability of the QuHE method, we conduct experiments on 100 uniformly sampled initial configurations for bandwidth, power, and computation frequencies. After optimization, the resulting objective values (shown in Fig. \ref{fig_optimality_gap_sample_dot}) range from a maximum of 10.95 (optimal) to a minimum of –20.77 (worst case).

To analyze the data, we calculate the proportion of samples yielding objective function values near the optimal and worst-case values. A solution is classified as ``very good'' if its objective function value is within $[10, 15]$, ``good'' if its objective function value is within $[5, 10]$, while a solution is deemed ``poor'' if its value is within $[-25, 0]$. From the results in Fig. \ref{fig_optimality_gap_sample_bar}, we know that very good solutions can be obtained at a 56\% chance, and at least good solutions can be obtained at an 88\% chance. Thus, we can get good approximation solutions at a very high probability, which shows the strong reliability of the proposed QuHE algorithm.
\begin{figure*}[htbp]
\centering
\subfigure[Convergence in Stage 1.]{
    \includegraphics[width=0.22\textwidth]{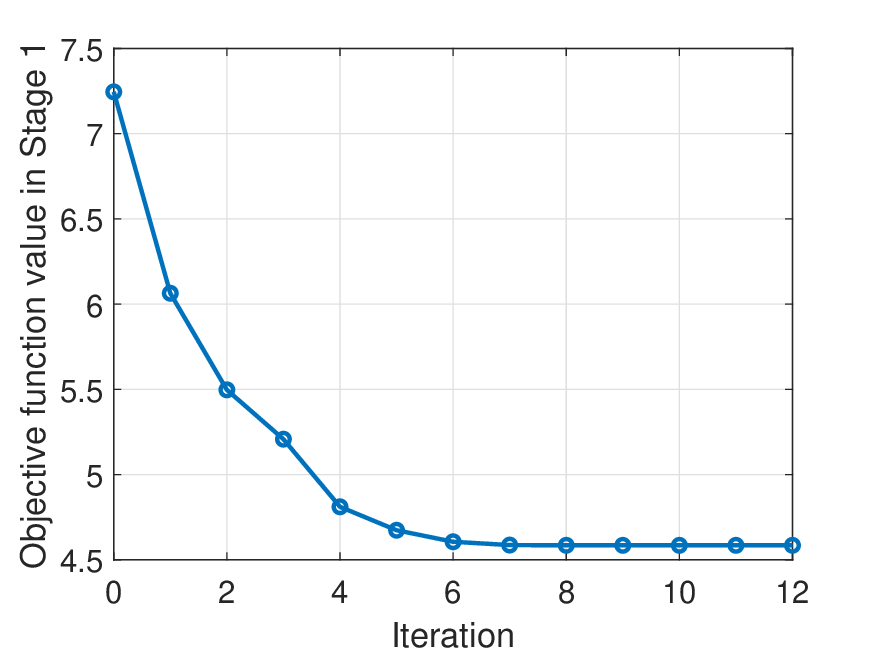}
    \label{fig_convergence_stage1}
}
\subfigure[Convergence in Stage 2.]{
    \includegraphics[width=0.22\textwidth]{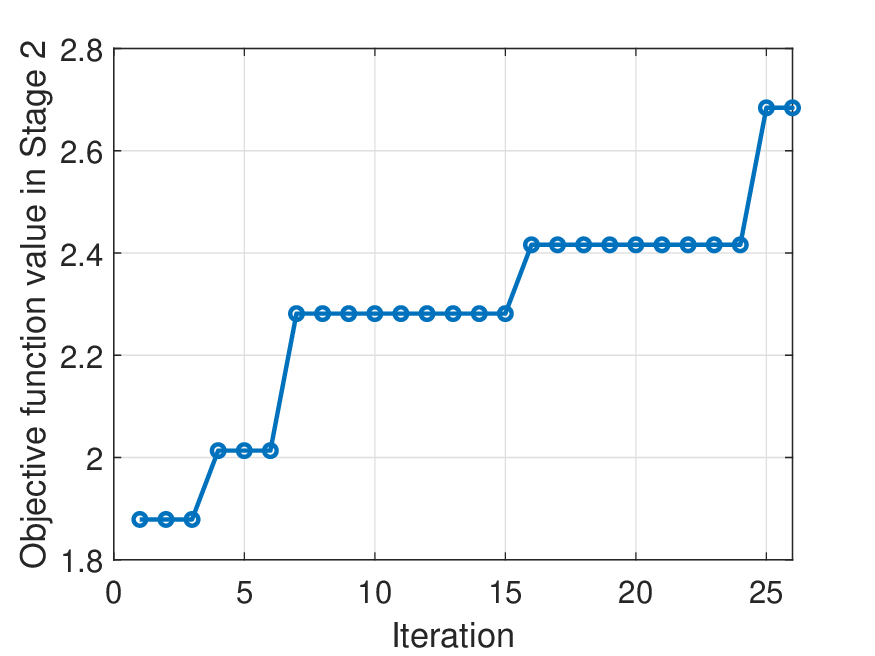}
    \label{fig_convergence_stage2}
}
\subfigure[Convergence in Stage 3.]{
    \includegraphics[width=0.22\textwidth]{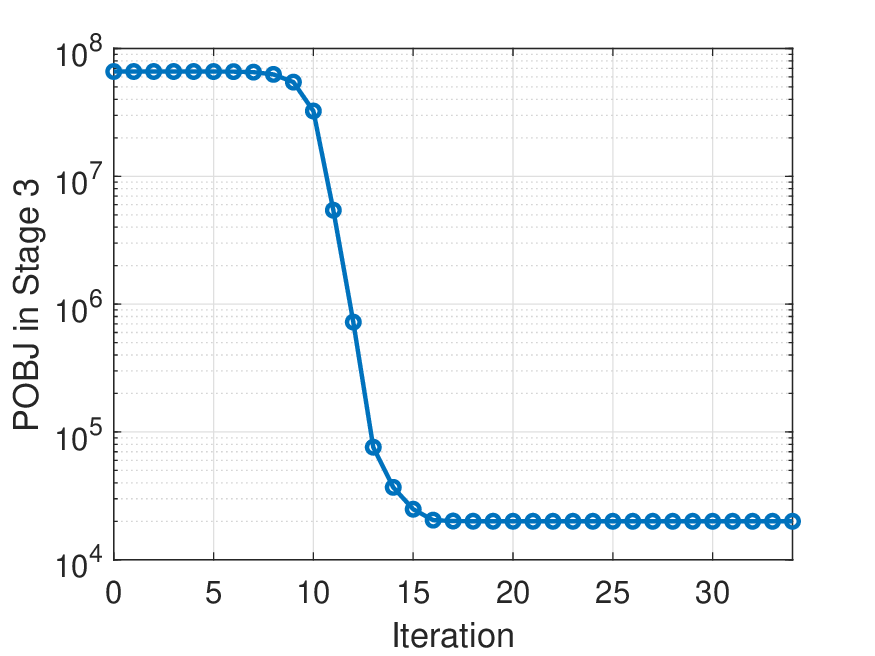}
    \label{fig_convergence_stage3}
}
\subfigure[Duality gap in Stage 3.]{
    \includegraphics[width=0.22\textwidth]{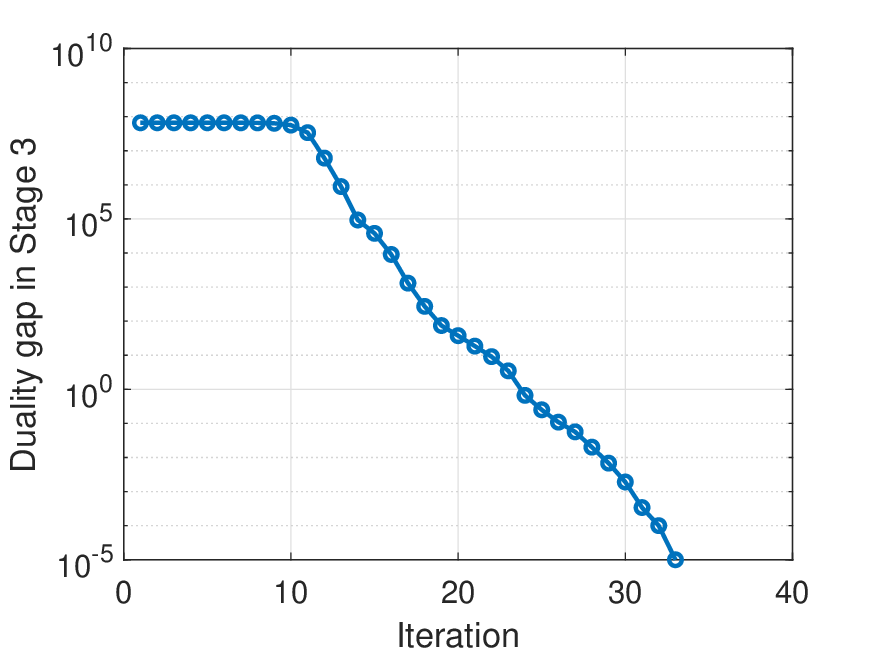}
    \label{fig_convergence_stage3_dualitygap}
}
\vspace{-6pt}
\caption{Convergence of the proposed QuHE algorithm in different stages. ``POBJ'' in Fig. \ref{fig_convergence_stage3} means the primal objective value in CVX. ``Duality gap'' in Fig. \ref{fig_convergence_stage3_dualitygap} is the gap between the primal and dual objectives in CVX.}
\label{fig_convergence_all}
\end{figure*}
\begin{figure*}[htbp]
\centering
\subfigure[Number of each stage call and running time in the QuHE method.]{
    \includegraphics[width=0.22\textwidth]{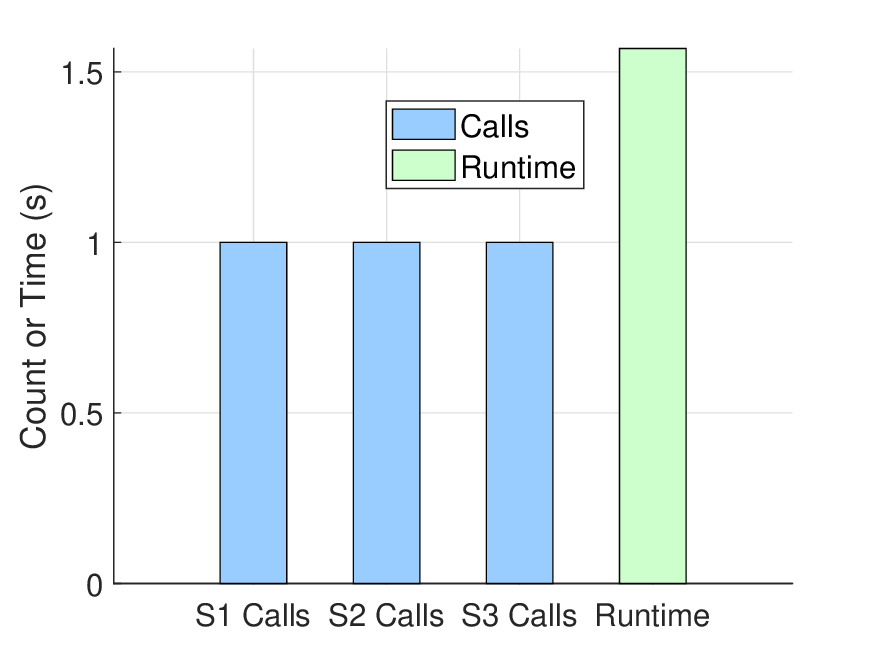}
    \label{fig_runtime_QuHE}
}
\subfigure[Running time of different methods in Stage 1.]{
    \includegraphics[width=0.22\textwidth]{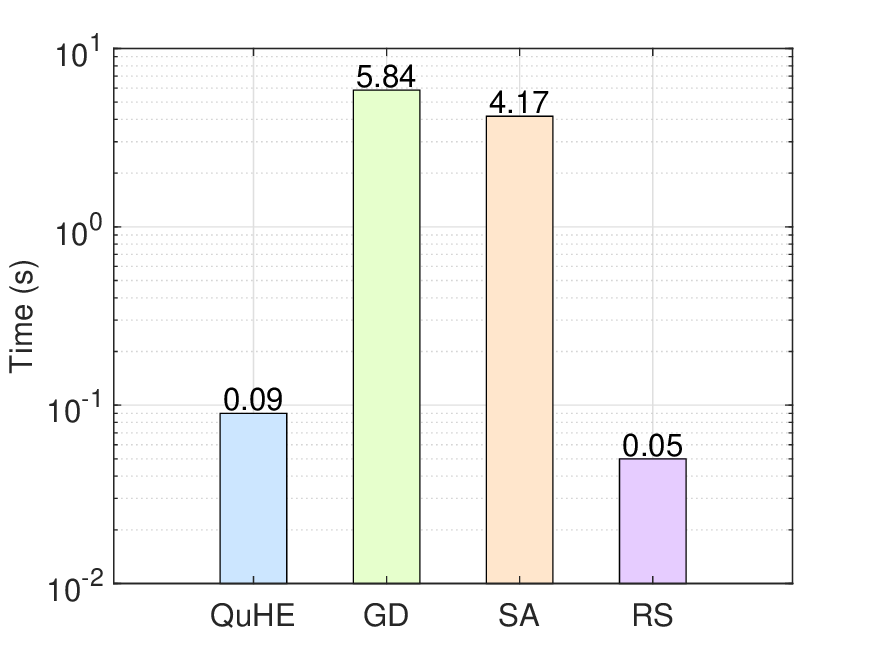}
    \label{fig_time_comparison_s1}
}
\subfigure[Performance comparison of different methods in Stage 1.]{
    \includegraphics[width=0.22\textwidth]{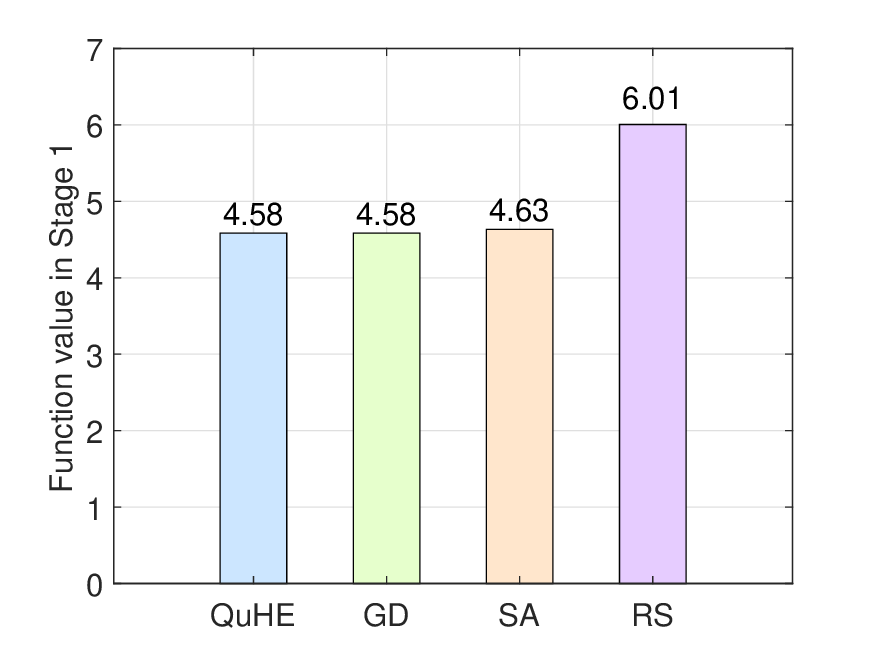}
    \label{fig_method_comparison_s1}
}
\subfigure[Performance comparison of different methods.]{
    \includegraphics[width=0.22\textwidth]{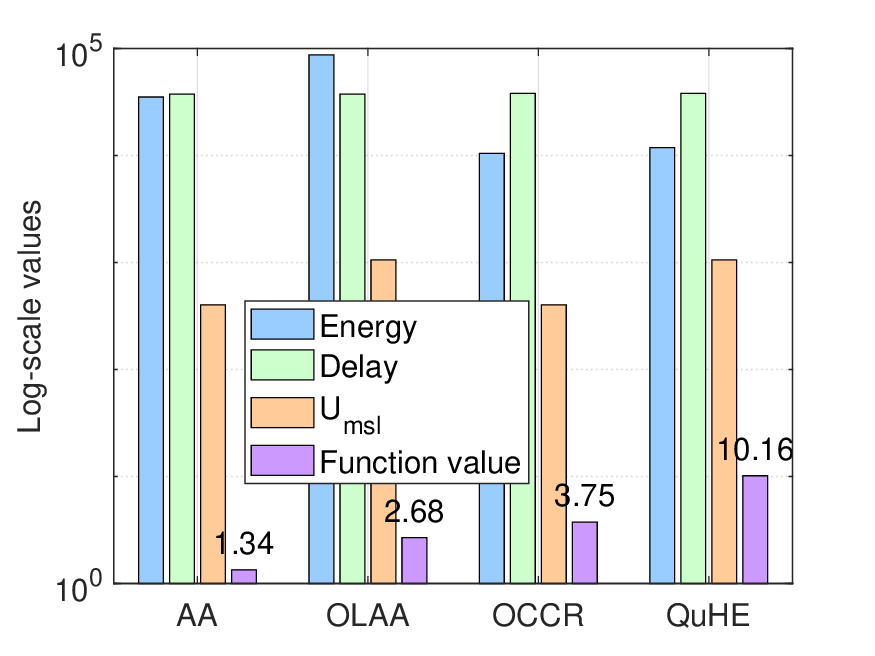}
    \label{fig_method_comparison}
}
\vspace{-6pt}
\caption{Running time and performance comparison of different methods. ``S1/2/3'' in Fig. \ref{fig_runtime_QuHE} mean Stage 1/2/3, respectively. ``GD'', ``SA'', and ``RS'' in Fig. \ref{fig_time_comparison_s1} and \ref{fig_method_comparison_s1} mean gradient descent, simulated annealing, and \vspace{-10pt} random selection, respectively. }
\label{fig_time_performance_comparison}
\end{figure*}
\begin{figure*}[htbp]
\centering
\subfigure[Performances under different total bandwidth settings.]{
    \includegraphics[width=0.22\textwidth]{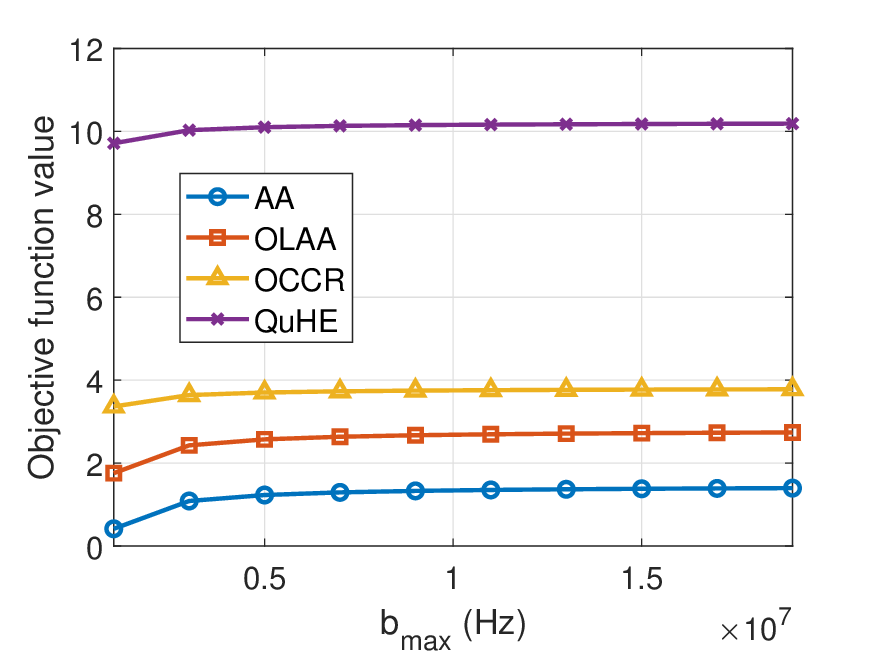}
    \label{fig_b_comparison}
}
\subfigure[Performances under different transmit power settings.]{
    \includegraphics[width=0.22\textwidth]{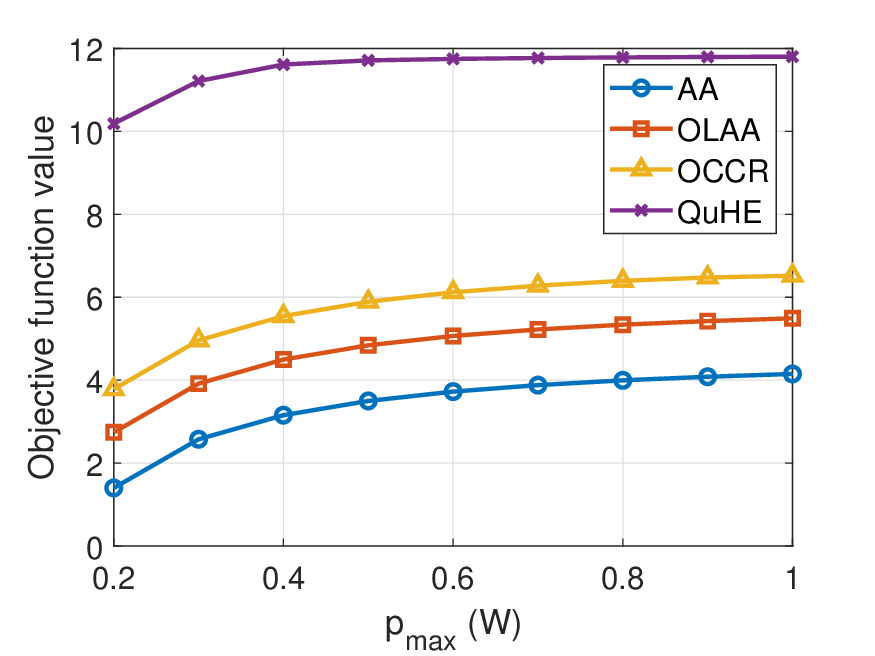}
    \label{fig_p_comparison}
}
\subfigure[Performances under different available computing resource settings at the client side.]{
    \includegraphics[width=0.22\textwidth]{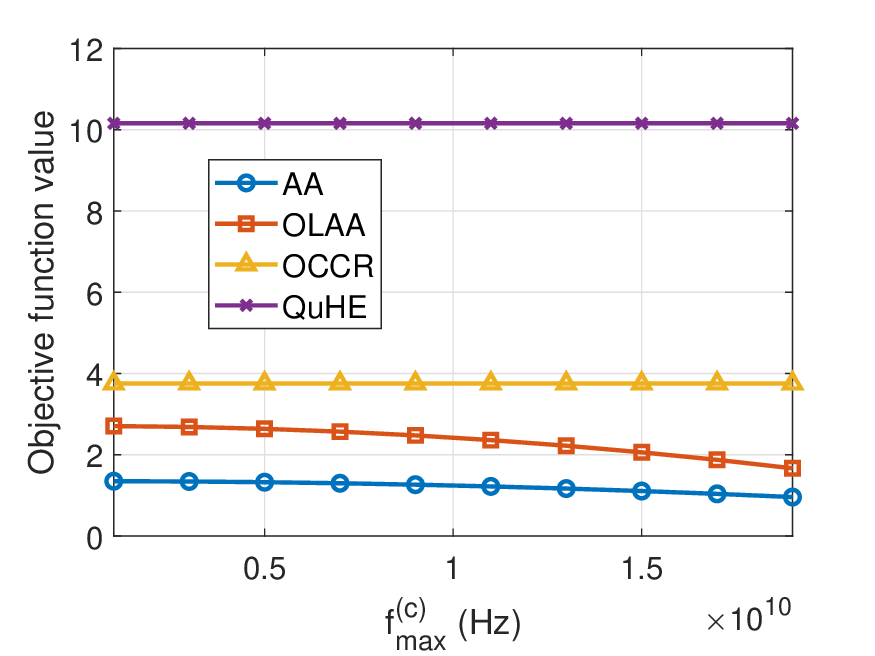}
    \label{fig_fc_comparison}
}
\subfigure[Performances under different available computing resource settings at the server side.]{
    \includegraphics[width=0.22\textwidth]{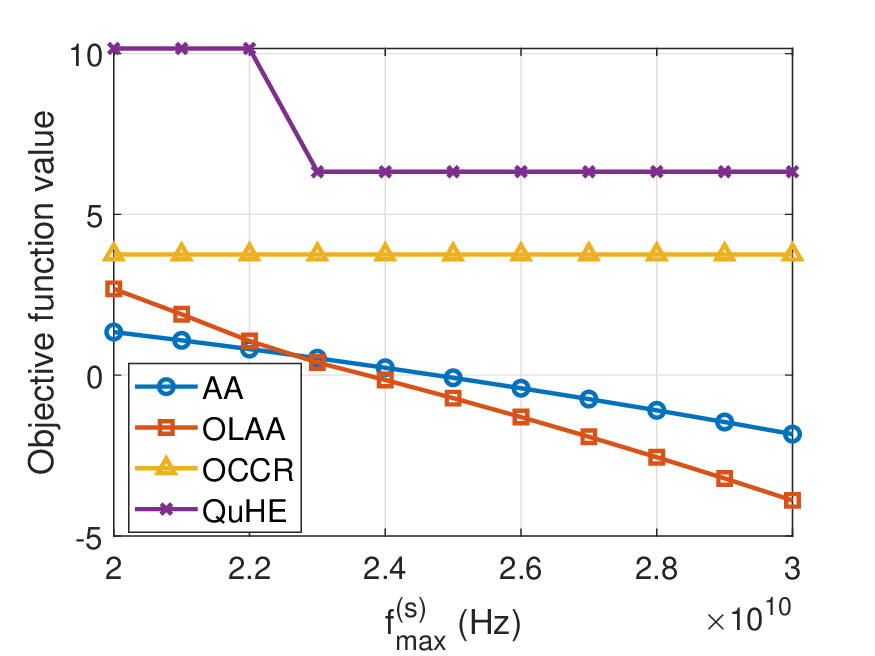}
    \label{fig_fs_comparison}
}
\vspace{-6pt}
\caption{Performance comparison of different methods under various computing and communication resource settings.}
\label{fig_compare_all}
\end{figure*}
\subsection{Convergence Analysis}
In Fig. \ref{fig_convergence_all}, we present the convergence of each stage in the proposed QuHE method. From the result, we know that our method converges within at most 34 iteration steps. Specifically, Stage 1 converges in 12 steps; Stage 2 converges in 26 steps; Stage 3 converges in 34 steps. The duality gap in Stage 3 achieves $10^{-5}$ at the 33-rd iteration. In Fig. \ref{fig_runtime_QuHE}, the number of each stage call and the related running time are given. We know that the proposed QuHE algorithm can converge in one call of each stage, and the total running time is 1.5 seconds, demonstrating the efficiency of the QuHE algorithm.

\subsection{Performance Analysis at Stage 1}
In Tables \ref{tab:phi_comparison} and \ref{tab:w_comparison}, we give the obtained optimal $\bm{\phi}$ and $\bm{w}$ values of different methods. The random selection method obtains more highest $\phi_n$, while the QuHE Stage 1 algorithm and gradient descent get more highest $w_l$. Although the gradient descent method achieves the same optimal solutions as the proposed QuHE Stage 1 method, the running time needed is much higher than that needed by the QuHE Stage 1 method in Fig. \ref{fig_time_comparison_s1}. In Fig. \ref{fig_method_comparison_s1}, the QuHE Stage 1 and gradient descent methods obtain the optimal objective function value. To conclude, the proposed QuHE Stage 1 method can achieve a better trade-off between the running time and the solution optimality than other baselines. 

\subsection{Performance Analysis of the Entire Process}
In Fig. \ref{fig_method_comparison}, we compare the performance of the AA, OLAA, OCCR, and QuHE methods based on energy consumption, system delay, minimum security level, and objective function value, assuming the optimal $U_{qkd}$ is obtained in \hbox{Stage 1}. The results show that QuHE and OCCR excel in energy efficiency, significantly outperforming AA and OLAA. In terms of system delay, all methods deliver comparable performance, with QuHE exhibiting a slightly higher delay. Regarding the security level, QuHE and OLAA achieve the highest scores, substantially surpassing AA and OCCR. Notably, QuHE stands out with the best overall objective function value, reflecting its ability to balance high security, energy efficiency, and low delay. Overall, QuHE consistently outperforms the other methods across all metrics.

\subsection{Performance Analysis under Various Computing and Communication Resource Settings}





Figure \ref{fig_compare_all} analyzes the performance of AA, OLAA, OCCR, and QuHE under varying resource settings.

Impact of $p^{(max)}_n$: Higher $p^{(max)}_n$ significantly improved all methods, with QuHE achieving the best results, showcasing superior power optimization.

Impact of $B_{total}$: Increases in $B_{total}$ had a marginal effect on AA and OLAA but yielded notable gains for QuHE and OCCR, with QuHE outperforming others.

Impact of $f^{(max)}_n$: Performance gains diminished as $f^{(max)}_n$ increased, though QuHE maintained the highest objective values despite rising energy consumption.

Impact of $f_{total}$: AA and OLAA struggled with increasing $f_{total}$, while OCCR and QuHE showed stability, with QuHE consistently leading.

Overall, QuHE demonstrated robust and superior performance across all scenarios, effectively optimizing resource allocation under diverse constraints.

\section{Conclusion}\label{sec_conclusion}
This paper introduces a novel framework that integrates QKD and HE into MEC systems, addressing the critical trade-off between QKD network utility, HE security levels, processing costs, and wireless transmission costs. By using QKD to securely distribute symmetric keys and HE for encrypted data processing, our proposed QuHE algorithm effectively optimizes the overall system performance. Theoretical and numerical analyses confirm the algorithm's reliability, efficiency, and superiority over baselines.

\section*{Acknowledgement}
This research is supported by the National Research Foundation, Singapore and Infocomm Media Development Authority under its Trust Tech Funding Initiative, Singapore Ministry of Education Academic Research Fund RG90/22, and Nanyang Technological University (NTU)-Wallenberg AI, Autonomous Systems and Software Program (WASP) Joint Project. Any opinions, findings and conclusions or recommendations expressed in this material are those of the author(s) and do not reflect the views of National Research Foundation, Singapore and Infocomm Media Development Authority.

\bibliographystyle{IEEEtran}
\bibliography{ref}

\end{document}